\documentclass[preprint,aps,showpacs,nofootinbib]{revtex4}

\usepackage{epsfig,amssymb,amsmath}

\def\bea{\begin{eqnarray}}
\def\eea{\end{eqnarray}}
\def\beq{\begin{equation}}
\def\eeq{\end{equation}}

\begin{document}
\draft
\tighten
\preprint{KIAS-P09042}
\title{\large \bf
    Thermal inflation and baryogenesis in heavy gravitino scenario}
\author{
    Kiwoon Choi\footnote{email: kchoi@muon.kaist.ac.kr}$^1$,
    Kwang Sik Jeong\footnote{email: ksjeong@kias.re.kr}$^2$,
    Wan-Il Park\footnote{email: wipark@muon.kaist.ac.kr}$^1$,
    Chang Sub Shin\footnote{email: csshin@muon.kaist.ac.kr}$^1$}
\affiliation{
    $^1$Department of Physics, KAIST, Daejeon 305-701, Korea \\
    $^2$School of Physics, Korea Institute for Advanced Study,
    Seoul 130-722, Korea}

\vspace{2cm}

\begin{abstract}

We present a thermal inflation model that incorporates the
Affleck-Dine leptogenesis in heavy gravitino/moduli scenario, which
solves the moduli-induced gravitino problem while producing a
correct amount of baryon asymmetry and relic dark matter density.
The model involves two singlet flat directions stabilized by
radiative corrections associated with supersymmetry breaking, one
direction that generates the Higgs $\mu$ and $B$ parameters, and the
other direction that generates the scale of spontaneous  lepton
number violation. The dark matter is provided by the lightest
flatino which might be identified as the axino if the model is
assumed to have a $U(1)_{PQ}$ symmetry to solve the strong CP
problem. We derive the conditions for the model to satisfy various
cosmological constraints coming from the Big-Bang nucleosynthesis
and the dark matter abundance.

\end{abstract}

\maketitle


\section{Introduction}

Low energy supersymmetry (SUSY) is one of the most plausible
candidates for physics beyond the Standard Model (SM) at the TeV
scale \cite{Nilles:1983ge}. In the context of supergravity, SUSY can
be spontaneously broken in a hidden sector while giving a vanishing
cosmological constant. In this framework, the gravitino mass is
related to the scale of hidden sector SUSY breaking as
$M_{\rm SB}\sim \sqrt{m_{3/2}M_{Pl}}$, where
$M_{Pl}=2.4\times 10^{18}$ GeV is the reduced Planck mass.
In low energy SUSY scenario, the hidden sector SUSY breaking is
transmitted to the supersymmetric standard model (SSM) to induce
soft terms providing sparticle masses of ${\cal O}(1)$ TeV.
Then, for certain range of $m_{3/2}$, gravitinos produced in the
early Universe decay after the Big-Bang nucleosynthesis, which would
destroy the successful prediction of the light element abundances
\cite{Pagels:1981ke}. This cosmological difficulty might be avoided
if the gravitino is relatively heavy, e.g.
$m_{3/2}\gtrsim {\cal O}(10)$ TeV, so that the decay occurs
before the nucleosynthesis.

Recent progress in string flux compactification \cite{Kachru:2003aw}
has provided a SUSY breaking scheme in which the hidden sector with
a SUSY breaking scale $M_{\rm SB}\sim \sqrt{m_{3/2}M_{Pl}}$ is
naturally sequestered from the SSM sector. String fluxes
generically produce a warped throat in compact internal space, and
then the SUSY breaking sector is stabilized at the IR end of throat.
On the other hand, the high scale gauge coupling unification
suggests that the SSM sector is located at the UV end, and thus is
sequestered from the SUSY breaking at the IR end of throat
\cite{Choi:2006bh,warped_sequestering}. In such case, the effective
contact interactions between the SUSY breaking hidden sector field
and the SSM fields are so suppressed \cite{rs} that the conventional
gravity mediation \cite{gravity.mediation} which would give a soft
mass of ${\cal O}(m_{3/2})$ does not contribute to the SSM soft
masses. As a result, the SSM soft masses can be much lighter than
the gravitino mass, and $m_{3/2}={\cal O}(10)$ TeV is a
natural outcome if one takes the SSM soft parameters to be near the
weak scale. Moreover, this framework can stabilize moduli either by
fluxes or by non-perturbative effects \cite{Kachru:2003aw}, and the
resulting moduli masses are typically much heavier than the
gravitino mass, $m_\phi\sim 8\pi^2 m_{3/2}$ for moduli stabilized by
non-perturbative effects and $m_\phi\gg 8\pi^2 m_{3/2}$ for moduli
stabilized by fluxes \cite{Choi:2004sx}.

Nonetheless, the above scenario of heavy gravitino/moduli is not yet
free from cosmological difficulty \cite{Endo:2006zj}. It has been
noticed that there still arises a problem, the moduli-induced
gravitino problem, due to that the gravitinos from moduli decays
produce too many neutralinos, which would overclose the Universe if
the neutralino is the stable lightest supersymmetric particle (LSP).
As another difficulty, in heavy gravitino scenario it is not
straightforward at all to get the Higgs $\mu$ and $B$ parameters
having a weak scale size, which would require a severe fine tuning
for the correct electroweak symmetry breaking. Recently, an
attractive solution to these problems has been proposed in
\cite{Nakamura:2008ey}, involving a singlet flat direction
stabilized by radiative corrections associated with SUSY breaking.
Once stabilized by radiative effects, the singlet flaton $S$ gets a
loop-suppressed SUSY breaking $F$-component, $F^S/S \sim
m_{3/2}/8\pi^2$, and then one can arrange the model to generate
$\mu\sim B\sim F^S/S$, which would result in a weak scale size of
$\mu$ and $B$ when $m_{3/2}={\cal O}(10)$ TeV. Such a scheme
can solve the moduli-induced gravitino problem also. Now the LSP of
the model is the fermionic component of $S$, the flatino, which is
much lighter than the MSSM neutralino. Then the neutralinos produced
by moduli/gravitino decays are not stable anymore, but decay into
light flatinos with a relic mass density not overclosing the Universe.
It is also natural to introduce a $U(1)_{PQ}$ symmetry spontaneously
broken by the vacuum value of $S$, and then the pseudoscalar
component of $S$ corresponds to the QCD axion solving the  strong CP
problem \cite{Kim:2008hd}, and the flatino can be identified as the
axino which has been proposed as a viable dark matter candidate in
\cite{Covi:1999ty}.

On the other hand, late time thermal inflation \cite{Lyth:1995hj} is
a natural consequence of generic flaton model, although its
possibility was not explored in \cite{Nakamura:2008ey}. Thermal
inflation also solves the moduli-induced gravitino problem by
diluting away the coherent oscillation of moduli. A potential
difficulty of thermal inflation is that any primordial baryon
asymmetry is also washed away by the late time entropy production,
so one needs a baryogenesis mechanism working after thermal
inflation is over\footnote{In gauge mediated SUSY breaking scenario,
the baryon asymmetry produced by the Affleck-Dine mechanism can
survive the dilution by thermal inflation if the initial amplitude
of the AD flat direction is large enough \cite{de Gouvea:1997tn}. In
such case, the curvature of the potential at a field value larger
than the messenger scale  is determined  by the small gravitino
mass, thus the oscillation of the flat direction begins much later
than the case of $m_{3/2} \gtrsim m_\mathrm{soft}$,
which would allow the generated baryon asymmetry large enough to
survive the late time dilution. In our case of heavy gravitino
scenario, this is not possible, so we need a baryogenesis after
thermal inflation is over.}. Recently, an interesting model of
thermal inflation incorporating the Affleck-Dine (AD) leptogenesis
\cite{Affleck:1984fy} has been proposed in the context of the
conventional gravity-mediated SUSY breaking scenario giving a weak
scale size of $m_{3/2}$
\cite{Jeong:2004hy,Felder:2007iz,Kim:2008yu}.

In this paper, we wish to examine the possibility of thermal
inflation incorporating a similar AD leptogenesis in heavy
gravitino/moduli scenario realized in the framework of sequestered
SUSY breaking. As we will see, a heavy gravitino mass requires
additional conditions for successful thermal inflation and AD
leptogenesis, e.g. the model should involve two singlet flat
directions both of which are stabilized by radiative effects, one
flat direction that generates the Higgs $\mu$ and $B$ parameters,
and the other flat direction that generates the scale of spontaneous
lepton number violation. We present a viable model satisfying
various cosmological constraints coming from the Big-Bang
nucleosynthesis and the dark matter abundance.

This paper is organized as follows. In section 2, we briefly discuss
the moduli-induced gravitino problem and also how to generate the
weak scale size of $\mu$ and $B$ with a radiatively stabilized
singlet flaton in heavy gravitino scenario. In section 3, we present
a specific model of thermal inflation and AD leptogenesis in heavy
gravitino scenario, and briefly analyze the phenomenological aspects
of the model. A detailed discussion of the cosmological aspects of
the model is provided in section 4, and section 5 is the conclusion.

\section{Low energy SUSY with heavy gravitino}
\label{Lsusy}

\subsection{Moduli-induced gravitino problem}

A gravitino mass much heavier than the SSM soft mass $m_{\rm soft}$
naturally appears in models with sequestered SUSY breaking. A SUSY
breaking at the tip of throat in string flux compactification
provides an attractive framework for sequestered SUSY breaking
\cite{warped_sequestering,Choi:2006bh}. Such string
compactifications include moduli that are stabilized by flux or
non-perturbative effect. The resulting moduli masses are comparable
to $8\pi^2 m_{3/2}$ for non-perturbative stabilization, and even
much heavier than $8\pi^2 m_{3/2}$ for flux stabilization
\cite{Choi:2004sx}. When SUSY breaking effects are taken into
account, those moduli $\phi$ develop a nonzero $F$-component given
by \bea \label{F-term-induced} \frac{F^\phi}{\phi} &\sim&
\frac{m^2_{3/2}}{m_\phi}, \eea where $\phi$ and $m_\phi$ denote the
modulus vacuum value and the modulus mass, respectively. Since
$m_\phi\gtrsim 8\pi^2 m_{3/2}$, the resulting moduli-mediated soft
masses do not dominate over the anomaly-mediated soft masses of
${\cal O}(m_{3/2}/8\pi^2)$\footnote{In fact, moduli-mediation
comparable to the anomaly mediation is phenomenologically desirable
since the pure anomaly mediation \cite{anomaly.mediation} suffers
from the tachyonic slepton problem in the minimal supersymmetric SM
\cite{dilaton.mediation,Choi:2008hn}.}
\cite{Choi:2004sx,Endo:2005uy,Choi:2005uz}. Thus, even when one
takes into account the moduli-mediated soft masses of ${\cal
O}(F^\phi/\phi)$, it still holds true the order of magnitude
relation $m_{\rm soft}\sim m_{3/2}/8\pi^2$, and the gravitino
generically has a heavy mass of ${\cal O}(8\pi^2)$ TeV when $m_{\rm
soft}$ is assumed to have a size of ${\cal O}(1)$ TeV.

Moduli are expected to have a misalignment of ${\cal O}(M_{Pl})$ in
the early Universe, and therefore their coherent oscillations would
soon dominate the energy density of the Universe. Since moduli
masses are much heavier than ${\cal O}(10)$ TeV in sequestered SUSY
breaking scenario, their decays take place well before the
nucleosynthesis. Nevertheless, moduli can cause a cosmological
problem unless the branching ratio of the moduli decay to gravitinos
is highly suppressed. Indeed, non-thermal gravitino production by
heavy moduli decay can lead to a severe over-abundance of dark
matter in the Universe \cite{Endo:2006zj}.

For a modulus $\phi$ with $m_\phi\gg m_{3/2}$, the number density of
gravitinos produced by the modulus decay is given by \bea
\frac{n_{3/2}}{s} &=& \frac{3}{2} \frac{\Gamma_{\phi\to\tilde
G\tilde G}}{\Gamma_{\phi\to{\rm MSSM}}} \frac{T^\phi_R}{m_\phi},
\eea where $s$ is the entropy density, $\Gamma_{\phi\to\tilde
G\tilde G}$ and $\Gamma_{\phi\to{\rm MSSM}}$ denote the decay width
to the gravitino pair and to the MSSM particles, respectively, and
$T^\phi_R$ is the reheating temperature for the modulus decays:
\bea
T^\phi_R &=& \Big(\frac{90}{\pi^2 g_\ast(T^\phi_R)}\Big)^{1/4}
\sqrt{\Gamma_\phi M_{Pl}} \,\simeq\,
0.2\Big(\frac{10}{g_\ast(T^\phi_R)}\Big)^{1/4}
\left(\frac{C_\phi}{10^{-1}}\right)^{1/2}
\left(\frac{m_\phi}{10^6{\rm GeV}}\right)^{3/2}{\rm GeV}, \eea where
$g_\ast(T)$ is the number of relativistic degrees of freedom at $T$,
and $\Gamma_\phi=C_\phi m^3_\phi/M^2_{Pl}$ is  the total modulus
decay width\footnote{For $\phi$ whose vacuum value determines the
gauge coupling constants, it decays mainly into the gauge bosons and
gauginos, and then  $C_\phi={\cal O}(10^{-1})$.}. Gravitinos
produced by modulus decay do not interact with others, and promptly
decay into the LSP. The decay temperature of the gravitino is given
by \bea T_{3/2} &\simeq& \Big(\frac{90}{\pi^2
g_\ast(T_{3/2})}\Big)^{1/4} \sqrt{\Gamma_{3/2} M_{Pl}}\, \sim\,
\left(\frac{m_{3/2}}{m_\phi}\right)^{3/2} T^\phi_R, \eea with
$\Gamma_{3/2}\sim 10^{-1}m^3_{3/2}/M^2_{Pl}$  being the total decay
width of the gravitino. Since $m_\phi\gg m_{3/2}$, $T_{3/2}$ is much
lower than $T^\phi_R$. Under the assumption of R-parity
conservation, if the LSP annihilation  is not efficient, each
gravitino will produce a stable LSP, yielding the LSP relic
abundance given by \bea \frac{\rho_\chi}{s} &\simeq& \frac{m_\chi
n_{3/2}}{s} \simeq 0.2 \,\frac{\rho_{\rm cr}}{s} \left(
\frac{\alpha}{1.4\times 10^{-5}} \frac{\Gamma_{\phi\to
\tilde{G}\tilde{G}}}{\Gamma_{\phi\to{\rm MSSM}}} \right), \eea where
$m_\chi$ denotes the LSP mass, \bea \alpha &=&
\left(\frac{C_\phi}{10^{-1}}\right)^{1/2}
\Big(\frac{10}{g_\ast(T^\phi_R)}\Big)^{1/4}
\left(\frac{m_\chi}{10^2\,{\rm GeV}}\right)
\left(\frac{m_\phi}{10^6\,{\rm GeV}}\right)^{1/2}, \eea and
$\rho_{\rm cr}/s\simeq 1.9\times10^{-9}$ GeV is the ratio of the
critical density to the entropy density in the present Universe.

Since $\rho_\chi/\rho_{\rm cr}$ in the present Universe cannot
exceed about 0.25, the branching ratio of the modulus decay into the
gravitino pair is bounded as \bea
\label{moduli-induced-gravitino-problem} \frac{\Gamma_{\phi\to\tilde
G\tilde G}}{\Gamma_{\phi\to{\rm MSSM}}} &\lesssim& 1.7\times
10^{-5}\left(\frac{10^{-1}}{C_\phi}\right)^{1/2}
\Big(\frac{g_\ast(T^\phi_R)}{10}\Big)^{1/4} \left(\frac{10^2\,{\rm
GeV}}{m_\chi}\right) \left(\frac{10^6\,{\rm
GeV}}{m_\phi}\right)^{1/2}. \eea In view of that $m_\phi\gtrsim
8\pi^2 m_{3/2}$ and $m_{3/2}={\cal O}(10)$ TeV in sequestered SUSY
breaking scenario,  $\Gamma_{\phi\to\tilde G\tilde G}$ should be
highly suppressed if $\chi$ corresponds to the MSSM neutralino
having a mass of ${\cal O}(10^2)$ GeV. However, it has been noticed
that there is no suppression by $m_{3/2}/m_\phi$ for the modulus
decay into the helicity $\pm 1/2$ components of the gravitino
\cite{Endo:2006zj}. As a consequence,  even in the limit $m_\phi\gg
m_{3/2}$, the branching ratio is simply given  by \bea
\frac{\Gamma_{\phi\to\tilde G\tilde G}}{\Gamma_{\phi\to{\rm
MSSM}}}\,\sim\, \frac{1}{4N_g}\,\sim\, 2\times 10^{-2}, \eea where
$N_g=12$ denotes the number of gauge bosons in the MSSM
\cite{Endo:2006zj}. This branching ratio exceeds the above bound by
several orders of magnitudes when $\chi$ is identified as the MSSM
neutralino, and this is the moduli-induced gravitino problem.

For a more careful analysis, one needs to include the effects of LSP
annihilation. Including such effect, we find  \bea
\frac{\rho_\chi}{s} &=& \frac{m_\chi n_\chi}{s} \simeq
\frac{1}{4}\left(\frac{90}{\pi^2 g_\ast(T_{3/2})}\right)^{1/2}
\frac{m_\chi}{\langle \sigma_{\rm ann} v_{\rm rel}\rangle
T_{3/2}M_{Pl}}, \eea where $\langle \sigma_{\rm ann}v_{\rm rel}\rangle$
is the thermal average of the annihilation cross section times the
relative velocity of $\chi$. If the branching ratio of
$\phi\rightarrow \tilde{G}\tilde{G}$ exceeds the bound
(\ref{moduli-induced-gravitino-problem}), in order not to overclose
the Universe,  the annihilation is required to be as efficient as
\bea \label{Annihilation} \langle \sigma_{\rm ann}v_{\rm rel}\rangle
&\geq& \frac{1.3\times 10^{-4}}{{\rm GeV}^{2}}
\left(\frac{10}{g_\ast(T_{3/2})}\right)^{1/4}
\left(\frac{m_\chi}{10^2\,{\rm GeV}}\right) \left(\frac{10^4\,{\rm
GeV}}{m_{3/2}}\right)^{3/2}. \eea On the other hand, even the Wino
LSP has an annihilation cross section $\langle \sigma_{\rm
ann}v_{\rm rel}\rangle \ll 10^{-4}\,{\rm GeV}^{-2}$, and thus it
appears to be difficult to avoid the moduli-induced gravitino
problem through the annihilation mechanism\footnote{In fact, in
other case that both $m_\phi$ and $m_{3/2}$ are of
${\cal O}(10)$ TeV, the resulting Wino LSP abundance can be
small enough not to overclose the Universe
\cite{moroi_randall,kane_etal}. However, in this case, the modulus
$\phi$ with $m_\phi\sim m_{3/2}$ generically has
$F^\phi/\phi\sim m_{3/2}$, and then the modulus-mediated
sfermion masses can be of ${\cal O}(10)$ TeV \cite{kane_etal}.
Here, we are focusing a different setup giving
$m_\phi\sim 8\pi^2 m_{3/2}$ (or heavier) with
$F^\phi/\phi\sim m_{3/2}/8\pi^2$ (or smaller), which is motivated by
sequestered SUSY breaking realized in string flux compactification.
In our case, even the Wino LSPs produced by moduli/gravitino decays
overclose the Universe by about two orders of magnitudes.}.

\subsection{Higgs $\mu$ and $B\mu$ term}

In heavy gravitino scenario, the MSSM Higgs $B$ parameter
generically receives a contribution of ${\cal O}(m_{3/2})$ from
SUGRA effect.
Unless cancelled by other contribution, such a large $B$ would make
it difficult to realize the correct electroweak symmetry breaking.
An attractive way to avoid this difficulty is to generate $\mu$ and
$B$ by a vacuum value of flat direction which is stabilized by
radiative corrections associated with SUSY breaking
\cite{Rattazzi:2000,Nakamura:2008ey}. Here we briefly discuss such a
scheme using the model of \cite{Nakamura:2008ey} as an example.

In addition to the canonical K\"ahler potential of the involved
fields, the model includes the following K\"ahler mixing and the
Yukawa interactions of the flaton field $S$: \bea \label{flaton1}
\Delta{\cal L}_{\rm int} &=& \int d^4\theta\,\kappa S^* S^\prime +
\int d^2\theta\,\Big[ y_H \Sigma H_u H_d + y_S \Sigma S S^\prime
\Big] + {\rm h.c.}, \eea which are allowed by the global $U(1)$
symmetry with the charge assignment:
$q(S)=q(S^\prime)=q(H_{u,d})=1,$ $q(\Sigma)=-2$. Then the direction
along $S\neq 0$ with $\Sigma=S^\prime=0$ is flat in the
supersymmetric limit. After integrating out the heavy fields
$\Sigma$ and $S^\prime$ under a large background value of $S$, one
obtains the effective theory of $S$ described by \bea
\label{GM-Higgs} \Delta{\cal L}_{\rm eff} &=& \int d^4\theta\left\{
Y^{\rm eff}_S S^*S + \left(\kappa_{\rm eff} \frac{S^*}{S}H_u H_d +
{\rm h.c.} \right) \right\}, \eea with $Y^{\rm eff}_S=Y_S(Q=|S|)$
and $\kappa_{\rm eff}=\kappa y_H/y_S$. Here $Q$ is the
renormalization scale, and $Y_S$ is the running wave function of
$S$. The potential is then determined by the running soft mass \bea
V &=& m^2_S(Q=|S|)|S|^2, \eea where $m^2_S=-F^IF^{J*}\partial_I
\partial_{\bar J}\ln Y_S$ for $F^I$ denoting generic SUSY-breaking
$F$-components in the model. The SUSY breaking $F^I$ include first
of all the $F$-component of the chiral compensator superfield
\bea C&=&C_0+\theta^2 F^C,\nonumber \eea as well as $F^S$ and the moduli
$F$-component $F^\phi$. For $S_0\equiv\langle S\rangle \gtrsim m_{3/2}$,
the equation of motion for $F^S$ reads \bea \frac{F^S}{S_0} &\simeq&
-F^I\partial_I \ln Y_S(Q=|S|). \eea It is quite possible that $Y_S$
is moduli-dependent at tree level, and then there will be a
contribution of ${\cal O}(F^\phi/\phi)$ to $F^S/S$. On the other
hand, due to the super-Weyl invariance in the compensator
formulation of SUGRA, the $C$-dependence of $Y_S$ arises only
through the RG running. As a result, the contribution from $F^C$ is
loop-suppressed, giving a contribution of ${\cal O}(F^C/8\pi^2 C_0)$
to $F^S/S_0$. In our case, $$\frac{F^C}{C_0}\,\simeq\, m_{3/2},\quad
\frac{F^\phi}{\phi}\, \sim\, \frac{m_{3/2}^2}{m_\phi}\, \lesssim\,
\frac{m_{3/2}}{8\pi^2},$$ and thus \bea \frac{F^S}{S_0} &=& {\cal
O}\left(\frac{m_{3/2}}{8\pi^2}\right). \eea Now the $\mu$ and $B\mu$
terms arise from the effective Higgs bilinear operator in the
effective  K\"ahler potential (\ref{GM-Higgs}): \bea \mu =
\kappa_{\rm eff} \left\{\left(\frac{F^S}{S_0}\right)^* + {\cal
O}\left(\frac{m_{3/2}}{8\pi^2}\right)\right\}, \quad B =
\frac{F^S}{S_0} + {\cal O}\left(\frac{m_{3/2}}{8\pi^2}\right), \eea
where the loop-suppressed contributions from $F^C$ are included in
${\cal O}(m_{3/2}/8\pi^2)$, together with the contributions from
$F^\phi$. Therefore, $\mu$ and $B$ generated by a radiatively
stabilized flat direction can be naturally of ${\cal
O}(m_{3/2}/8\pi^2)$, so be the weak scale when $m_{3/2}={\cal
O}(10)\,{\rm TeV}$.

\subsection{Axino dark matter and thermal inflation}

In heavy gravitino scenario, the addition of the singlet flaton $S$
has important implications for cosmology. Using the relation
$Y^{\rm eff}_S=Y_S(Q=|S|)$ and the stationary condition
$m^2_S(Q=S_0)\simeq 0$ for $S_0\gtrsim m_{3/2}$ in the effective
theory (\ref{GM-Higgs}), one can find that the radial flaton and the
flatino\footnote{This flatino can be called also the axino which
would be the right name if the global symmetry $U(1)_S$ is good
enough to be a $U(1)_{PQ}$ symmetry solving the strong CP problem
\cite{Kim:2008hd}.} acquire SUSY breaking masses \bea
\label{spectrum} m^2_{\sigma_S} &\simeq& \left.\frac{d
m^2_S(Q)}{d\ln Q}\right|_{Q=S_0} = {\cal
O}\left(\frac{m^2_{3/2}}{(8\pi^2)^3}\right),
\nonumber \\
m_{\psi_S} &\simeq& \left.\frac{y^2_S(Q)}{16\pi^2}A_S(Q)
\right|_{Q=S_0} = {\cal
O}\left(y^2_S\frac{m_{3/2}}{(8\pi^2)^2}\right), \eea where
\bea S&=&(S_0
+ \sigma_S/\sqrt 2)e^{i a_S/\sqrt 2 S_0}+\sqrt 2 \theta \psi_S
+\theta^2 F^S,\nonumber \eea
and $A_S$ is the soft A-parameter associated with the Yukawa
coupling $y_S$ in (\ref{flaton1}). The angular flaton ($=$ axion)
$a_S$ remains massless before the explicit breaking of $U(1)_S$ is
taken into account. As having a light mass suppressed by a loop
factor relative to the MSSM sparticle masses, the flatino is the LSP
and becomes a dark matter of the Universe under the usual assumption
of R-parity conservation \cite{Covi:1999ty,Chun:2000jx}.
Such a light dark matter is cosmologically favorable since the
moduli-induced gravitino problem would be considerably alleviated as
can be seen in (\ref{moduli-induced-gravitino-problem}).
This point has been used in \cite{Nakamura:2008ey} to solve the
moduli-induced gravitino problem by assuming
$m_{\psi_S}\sim 10^2$ MeV under the additional assumption
that there is no thermal inflation triggered by the flaton $S$.

However, since the flaton couples to the thermal bath in the early
Universe through the Yukawa interaction which is responsible for its
stabilization, it is a more plausible possibility that $S$ is hold
at the origin, $S=0$, by its thermal mass in the early Universe.
In that case, the Universe experiences a short period of thermal
inflation \cite{Lyth:1995hj,Jeong:2004hy} driven by the flaton
potential energy $V_0 \sim m^2_{\sigma_S} S^2_0$ around the origin.
After thermal inflation, $S$ would start to roll down towards its
true minimum $S_0$ and oscillate around the true minimum with an
amplitude of ${\cal O}(S_0)$. As it produces a tremendous amount of
entropy, thermal inflation practically dilutes away all the unwanted
relics, including the coherent oscillation of heavy moduli causing
the moduli-induced gravitino problem.

Thermal inflation is a natural consequence of the flaton model
providing a weak scale size of $\mu$ and $B$ in heavy gravitino
scenario, and immediately solves the moduli-induced gravitino
problem. However, there are certain cosmological constraints which
require a considerable extension of the model. After thermal
inflation, the Universe is reheated by the decays of the oscillating
radial flaton $\sigma_S$. If the axion $a_S$ is stable, the axion
energy density produced by the decays of $\sigma_S$ is bounded by
the Big-Bang nucleosynthesis, which requires that $\sigma_S$ decays
dominantly into the SM particles \cite{Choi:1996vz}. In the model
under consideration, to generate a weak scale size of $B$, it is
designed that $S$ couples to the operator $H_uH_d$ in the low energy
effective action (\ref{GM-Higgs}) through the combination $S^*/S$ at
tree level. With this structure of the model, the coupling of
$\sigma_S$ to $H_uH_d$ is cancelled at tree level, so its strength is
loop-suppressed. As the decays of $\sigma_S$ to the SM particles
are mediated by the coupling to $H_uH_d$, this results in that
$\sigma_S$ decays dominantly to the axions, so an overproduction of
axions which would be in conflict with the Big-Bang nucleosynthesis
\cite{Choi:1996vz}.

Another difficulty to be overcome is that thermal inflation dilutes
any pre-existing baryon asymmetry. Thus one needs to introduce a
baryogenesis mechanism which works after thermal inflation is over.
Recently, it has been noted that the observed baryon asymmetry can
be generated via the Affleck-Dine (AD) leptogenesis after thermal
inflation. The AD mechanism \cite{Affleck:1984fy} can be implemented
by a flat direction carrying nonzero lepton-number. A particularly
interesting candidate for such a flat direction is the MSSM $LH_u$
as it can have a large nonzero value during the period of $\mu=0$,
then rolls back to the origin after a nonzero $\mu$ is induced by
the flaton vacuum value. Here, $L$ denotes the lepton doublet
superfield, and we do not specify the generation structure for
simplicity. The dynamics of $LH_u$ is determined also by the dim $=$
5 neutrino mass operator: \bea \label{Lifting-AD} \Delta W_\nu &=&
\frac{LH_uLH_u}{M_\nu}, \eea which might be generated by the seesaw
mechanism \cite{seesaw}. Then, in order for the AD leptogenesis to
work, the $A$-type soft parameter associated with the neutrino mass
operator should satisfy \cite{Jeong:2004hy,Kim:2008yu} \bea
\label{AD-condition} |A_\nu|^2 &<& 6(m^2_{\tilde{L}} + m^2_{H_u}
+|\mu|^2), \eea where $m_i$ denote the soft scalar masses.
Otherwise, $LH_u$ is trapped at a meta stable minimum with $LH_u\neq
0$ even after $\mu\neq 0$ is generated. In heavy gravitino scenario,
this is a nontrivial requirement as $F^C$ of the SUGRA compensator
generically gives a contribution of ${\cal O}(m_{3/2})$ to $A_\nu$.

As we will see in the next section, all these difficulties of
thermal inflation in heavy gravitino scenario can be naturally
solved by introducing an additional singlet flat direction $X$.
Thermal inflation can successfully incorporate the AD leptogenesis
if $X$ is stabilized also by radiative effects associated with SUSY
breaking, and its vacuum value gives the heavy right-handed neutrino
mass inducing the neutrino mass operator $\Delta W_\nu$ via the
seesaw mechanism. With this additional flaton $X$, the
overproduction of axions is naturally avoided since now both
flatons, $S$ and $X$, have an unsuppressed coupling to $H_uH_d$,
with which the flatons decay dominantly into the SM particles.

\section{The model}

To resolve the difficulties noticed in the previous section, we
introduce an additional singlet flaton $X$ together with the
right-handed neutrino $N$ generating the neutrino mass operator
$\Delta W_\nu$ via the seesaw mechanism, and an extra matter pair
$\Psi,\Psi^c$ providing a Yukawa coupling to stabilize the original
flaton $S$. The relevant part of the model is given by \bea
\label{UV-theory} \Delta{\cal L}_{\rm int} &=& \int d^4\theta\,
\kappa S^* S^\prime +\int d^2\theta\Big[\,y_H \Sigma H_u H_d + y_X X
S^\prime\Sigma \,\Big]
\nonumber \\
&& + \int d^2\theta\Big[\, y_N N L H_u + \frac{1}{2}y^\prime_N X N
N+ \frac{1}{2}(\lambda_{S} S +\lambda_{S^\prime}S^\prime)\Psi\Psi^c
\,\Big] + {\rm h.c.}, \eea where we have imposed two global $U(1)$
symmetries, $U(1)_S$ and $U(1)_X$,  with the following charge
assignments \bea U(1)_S &:&\,\, (2,2,0,-2,0,-1,2,-2),
\nonumber \\
U(1)_X &:&\,\, (0,0,2,-2,-1,0,2,0) \nonumber \eea for the fields
$S,S^\prime,X,\Sigma,N,L,H_uH_d$ and $\Psi\Psi^c$, respectively.
Obviously both $X$ and $S$ correspond to flat directions when
$\Sigma, S^\prime, N, \Psi$ and $\Psi^c$ are all frozen at the
origin. Here the extra matter fields $\Psi,\Psi^c$ can be either
gauge-singlet or gauge-charged. In case of gauge-charged
$\Psi,\Psi^c$, our model can be considered as a simple
generalization of some models discussed in \cite{Everett:2008qy},
incorporating thermal inflation and AD leptogenesis.

Assuming that both $X$ and $S$ get  large vacuum values, one can
integrate out the heavy $\Sigma, S^\prime, N, \Psi$ and $\Psi^c$.
The resulting effective theory can be written as \bea \label{Model}
{\cal L}_{\rm eff} &=& {\cal L}_{\rm MSSM} + \int d^4\theta \left[\,
Y^{\rm eff}_X X^* X + Y_S^{\rm eff} S^*S+ \left(\,\hat\kappa
\frac{S^*}{X} H_u H_d+{\rm h.c.}\,\right) \,\right]
\nonumber \\
&& + \left(\,\int
d^2\theta\,\frac{1}{2}\lambda_\nu\frac{LH_uLH_u}{X} + {\rm h.c.}\,
\right), \eea where $\hat\kappa = \kappa y_H/y_X$ and
$\lambda_\nu=y^2_N/y^\prime_N$. Here the effective wave function
coefficient of $\varphi=S,X$ is given by $Y^{\rm
eff}_{\varphi}=Y_{\varphi}(Q=|\varphi|)$, where the running
wavefunction coefficient  $Y_{\varphi}(Q)$ can be computed from the
underlying theory (\ref{UV-theory}).

\subsection{Flaton stabilization}

For the underlying theory given by (\ref{UV-theory}), the mixing
between the two flatons, $X$ and $S$, in the effective potential is
negligible. The flaton potential is again induced by radiative
effects associated with SUSY breaking, thus  can be written in terms
of the running soft masses: \bea \label{V-XS} V &\simeq&
\sum_{\varphi=X,S} m^2_\varphi(Q=|\varphi|)|\varphi|^2, \eea where
\bea m^2_\varphi(Q) &=& -F^IF^{J*}\partial_I\partial_{\bar J}\ln
Y_\varphi(Q). \eea Here we consider the case that $m^2_\varphi(Q)$
is driven to be negative at certain low scales by the associated
Yukawa interaction. Then the flaton field is stabilized at
$\varphi=\varphi_0$ satisfying \bea m^2_\varphi(Q=\varphi_0)
&\simeq& 0, \eea and one can find the flaton $F$-term is given by \bea
\label{F-EOM} \frac{F^\varphi}{\varphi} &\simeq& -F^I\partial_I
\ln Y_\varphi(Q=|\varphi|), \eea where $F^I$ denote the SUSY
breaking $F$-components in the model, including $F^C\simeq m_{3/2}$
and the moduli $F$-component $F^\phi\sim m_{3/2}^2/m_\phi\lesssim
m_{3/2}/8\pi^2$. Again, for both $S$ and $X$, the resulting
$F^\varphi/\varphi_0$ is of the order of the MSSM soft mass $m_{\rm
soft}\sim m_{3/2}/8\pi^2$ as desired.

Around the minimum of potential, the radial flaton $\sigma_\varphi$
and the flatino $\psi_\varphi$ acquire SUSY breaking masses
\bea
\label{SX-masses}
m^2_{\sigma_\varphi} &\simeq& \left.\frac{d
m^2_\varphi(Q)}{d\ln Q}\right|_{Q=\varphi_0} = {\cal
O}\left(\frac{m^2_{\rm soft}}{8\pi^2}\right),
\nonumber \\
m_{\psi_\varphi} &\simeq& \sum_{ij}\left.
\frac{y^2_{\varphi ij}(Q)}{16\pi^2} A_{\varphi ij}(Q)\right|_{Q=\varphi_0}
= {\cal O}\left(\frac{m_{\rm soft}}{8\pi^2}\right), \eea where
$y_{\varphi ij}$ are the Yukawa couplings, $A_{\varphi ij}$ are the soft
$A$-parameters in the canonical basis. Here we are using the
parametrization
\bea
\varphi &=&
\left(\varphi_0+\frac{\sigma_{\varphi}}{\sqrt{2}}\right)
e^{ia_{\varphi}/\sqrt{2}\varphi_0}+\sqrt{2}\theta
\psi_{\varphi}+\theta^2F^\varphi,
\nonumber
\eea
for $\varphi=X,S$,  and also the fact that the mixing between $X$
and $S$ is negligible. The axion component $a_\varphi$ would remain
massless unless the associated global $U(1)$ symmetry is broken by
higher dimensional operators or by non-perturbative effect. Because
the flatinos have masses of ${\cal O}(m_{\rm soft}/8\pi^2)$, the
lightest flatino will constitute the dark matter under the
assumption of R-parity conservation.

\subsection{Higgs $\mu$ and $B$, and the effective potential of $LH_u$}

In the model in consideration, the Higgs $\mu$ and $B$ parameters
are induced through the coupling of $H_uH_d$ to the flaton field
combination $S^*/X$ in the effective lagrangian (\ref{Model}).
Including the potentially possible contribution from the moduli
$F$-component $F^\phi/\phi\sim m_{3/2}^2/m_\phi\lesssim
m_{3/2}/8\pi^2$, the resulting $\mu$ and $B$ are given by \bea
\label{b-mu} \mu &=& \hat\kappa \frac{S_0}{X_0}
\left\{\left(\frac{F^S}{S_0}\right)^* + {\cal O}(m_{\rm soft})
\right\},
\nonumber \\
B &=& \frac{F^X}{X_0} + {\cal O}(m_{\rm soft}), \eea where the
contributions from $F^\phi$ are in ${\cal O}(m_{\rm soft})$.
One then finds both  $\mu$ and $B$ have a desirable size, i.e.
${\cal O}(m_{\rm soft})$ when $\hat\kappa S_0\sim X_0$, which is
rather a natural possibility.

In the presence of the neutrino mass operator $\Delta
W_\nu=\lambda_\nu LH_uLH_u/2X$, the effective potential for the $LH_u$
flat direction is given by
\bea V_{LH_u} &=& \frac{1}{2}(m^2_{\tilde
L}+m^2_{H_u}+|\mu|^2)|\ell\,|^2 + \left( \frac{A_\nu
\lambda_\nu}{8}\frac{\ell^4}{X_0} + {\rm c.c.} \right) +
\frac{|\lambda_\nu|^2}{4} \frac{|\ell\,|^6}{X^2_0}, \eea where
$\ell$ is the field variable parameterizing the $LH_u$ flat
direction, and $A_\nu$ is the soft $A$-parameter for $\Delta W_\nu$,
which is given by \bea \label{A-nu} A_\nu &=& \frac{F^X}{X_0} +
{\cal O}(m_{\rm soft}). \eea During the period before a nonzero
$\mu$ is induced by the flaton vacuum value, if
$m^2_{\tilde{L}}+m^2_{H_u}<0$, which will be assumed in the
following, $\ell$ is stabilized at
\bea \ell_0 &\sim& \sqrt{|A_\nu X_0/\lambda_\nu|}.\eea If the condition
(\ref{AD-condition}) is satisfied, which is easily done in our model,
$\ell$ is correctly rolling back  to the origin after implementing
the AD leptogenesis.
Note that it is crucial for the AD leptogenesis to work that $A_\nu$
is of the order of $m_{\rm soft}\sim m_{3/2}/8\pi^2$, which is
achieved in our model by generating the scale of lepton number
violation by a radiatively stabilized  vacuum value $X_0$.

\subsection{Flaton decay}

In this subsection, we examine the decay of the radial flaton
$\sigma_\varphi$  $(\varphi=S,T)$. Since
$m_{\sigma_\varphi}={\cal O}(m_{\rm soft}/4\pi)$, these radial flatons
are kinematically forbidden to decay into the MSSM sparticles,
the gravitino, or the moduli. The axion components $a_\varphi$ can
acquire a small mass from non-perturbative effect or higher dimensional
operator breaking $U(1)_\varphi$, but they will be taken to be almost
massless in the following. In the low energy effective theory, the decays
of $\sigma_\varphi$ into axions or flatinos  are mediated by the
interactions \bea {\cal L}_{\rm int} &=&
\frac{\sigma_\varphi}{2\sqrt{2}\varphi_0} (\partial^\mu
a_\varphi)(\partial_\mu a_\varphi) + \lambda_{\psi_\varphi}
\frac{m_{\psi_\varphi}}{2\sqrt{2}\varphi_0}
\sigma_\varphi\psi_\varphi\psi_\varphi + {\rm h.c.}, \eea where
$\lambda_{\psi_\varphi}$ is determined by the running
$A$-parameters, $A_{\varphi ij}$, associated with the Yukawa
couplings, $y_{\varphi ij}$, responsible for the flaton
stabilization: \bea \lambda_{\psi_\varphi} &\simeq&
\frac{1}{\sum_{kl}y^2_{\varphi kl}A_{\varphi kl}}\sum_{ij}\left.
\left( \frac{d y^2_{\varphi ij}}{d \ln Q} A_{\varphi ij} +
y^2_{\varphi ij}\frac{d A_{\varphi ij}}{d\ln Q} \right)
\right|_{Q=\varphi_0}, \eea where $i,j$ and $k,l$ run over the
fields having a nonzero Yukawa coupling with the flaton
$\varphi=X,S$. One then finds that generically
$\lambda_{\psi_\varphi}$ has a value of ${\cal O}(10^{-1})$ or less.
Then the decay rates are estimated as \bea
\label{Flaton-decay-rate}
\Gamma_{\sigma_\varphi\to a_\varphi a_\varphi} &=&
\frac{1}{64\pi}\frac{m^3_{\sigma_\varphi}}{\varphi^2_0},
\nonumber \\
\Gamma_{\sigma_\varphi\to\psi_\varphi\psi_\varphi} &=&
\frac{\lambda^2_{\psi_\varphi}}{32\pi} \frac{m^2_{\psi_\varphi}
m_{\sigma_\varphi}}{\varphi^2_0}, \eea where we have taken  $\hat
\kappa S_0\sim X_0$.

The radial flatons decay also to the SM particles, mainly through
the effective interactions \bea {\cal L}_{\rm int} &=&
\frac{1}{\sqrt
2}\left(\frac{\sigma_X}{X_0}-\frac{\sigma_S}{S_0}\right)
\left(|\mu|^2 |H^0_u|^2+|\mu|^2 |H^0_d|^2 - B\mu H^0_uH^0_d \right)
+ {\rm h.c.}, \eea which would induce a mass mixing between
$\sigma_\varphi$ and the neutral Higgs fields after the electroweak
symmetry breaking. The flaton-Higgs mixing then allows
$\sigma_\varphi$ to decay directly into the SM fermions, which is
dominated by the bottom quark channel: \bea
\Gamma_{\sigma_\varphi\to b\bar b}\, \sim\, \frac{3}{4\pi} \left(1 -
\frac{|B|^2}{m^2_A} \right)^2
\left(1-\frac{4m^2_b}{m^2_{\sigma_\varphi}} \right)^{3/2}
\left(\frac{|\mu|^2}{m^2_h}\right)^2 \frac{m^2_b
m_{\sigma_\varphi}}{\varphi^2_0}, \eea where $b$, $h$, and $A$ are
the bottom quark, the lightest neutral CP even Higgs boson, and the
neutral CP odd Higgs boson, respectively.

After thermal inflation is over, the Universe is dominated by the
energy density of coherently oscillating radial flaton
$\sigma_{\varphi}$. If $\sigma_\varphi$ decays dominantly to axions,
it would be in conflict with the Big-Bang nucleosynthesis
\cite{Choi:1996vz}. With the above results, the branching ratio to
the decay into axions is given by \bea
\frac{\Gamma_{\sigma_\varphi\rightarrow a_\varphi
a_\varphi}}{\Gamma_{\sigma_\varphi\rightarrow b\bar{b}}}&\sim&
\frac{1}{48}\left(\frac{m_{\sigma_\varphi}}{m_b}\right)^2
\left(\frac{m_h^2}{|\mu|^2}\right)^2,
\eea which can be easily smaller than ${\cal O}(10^{-1})$ to
satisfy the bound from the Big-Bang nucleosynthesis.

Meanwhile, assuming $X_0>S_0$, the following interaction between two
flatinos is induced by the $X$ dependence of $Y^{\rm eff}_S$: \bea
\label{Singlino-interaction}
{\cal L}_{\rm flatino} &=& \frac{\lambda_\psi}{\sqrt 2 X_0} \psi_X
\sigma^\mu \bar \psi_S (\partial_\mu a_S) + {\rm h.c.}, \eea where
$\lambda_\psi=\langle X \partial_X \ln Y^{\rm eff}_S \rangle$. Then
the heavier flatino $\psi_1$ decays into the lighter flatino
$\psi_0$  plus an axion with the decay rate \bea
\label{flatinodecay}
\Gamma_{\psi_1\to\psi_0} &\simeq& \frac{|\lambda_\psi|^2}{32\pi}
\left(1-\frac{m^2_{\psi_0}}{m^2_{\psi_1}}\right)^3
\frac{m^3_{\psi_1}}{X^2_0}. \eea
Since $\lambda_\psi$ is induced at higher than two-loop level, its
magnitude is suppressed as $|\lambda_\psi|<{\cal O}(1/(8\pi^2)^2)$,
so that $\psi_1$ decays with a long lifetime
while producing an axion together with the axino dark matter.
Cosmological implication of such a late decay of the heavier flatino
will be discussed in the next section.

\section{Cosmology}

The singlet flaton fields in our model allow a natural realization
of thermal inflation and baryogenesis, and provide light flatinos
one of which is the LSP. We assume, as the boundary condition of our
cosmology, that radiation is dominant before the commencement of
moduli coherent oscillation, and the temperature was high enough to
hold all fields except moduli near the origin. As temperature drops
down, the flaton field $\varphi$ starts to roll down towards its
true minimum $\varphi=\varphi_0$ at the critical temperature
$T_\varphi$: \bea T_\varphi &=& \frac{\hat m_\varphi}{\beta_\varphi}
\quad (\varphi=X,S),\eea where $\hat m^2_\varphi=-m_\varphi^2(Q\sim
m_{\rm soft})>0$ is the soft scalar mass squared of $\varphi$
renormalized at $Q \sim m_\mathrm{soft}$, and $\beta_\varphi$
determines how strongly $\varphi$ couples to thermal bath
\cite{Comelli:1996vm} \bea \beta^2_\varphi &=& \frac{1}{8}
\sum_{ij}|y_{\varphi ij}|^2. \eea Before the Higgs $\mu$ term is
induced by the flaton vacuum value, $\ell$ parameterizing the $LH_u$
flat direction is also thermally trapped at the origin, and its
critical temperature is given by \bea T_\ell &=& \frac{\hat
m_\ell}{\beta_\ell}, \eea where $\hat m^2_\ell=-(m^2_{\tilde
L}+m^2_{H_u})/2>0$, and $\beta_\ell$ is given by \bea \beta^2_\ell
&=& \frac{1}{8}\Big( \sum_{ij}|y_{L ij}|^2 +  4 \sum_a C^a_2(L)
g^2_a \Big) + (L\leftrightarrow H_u), \eea where $C^a_2(\Phi)$
denotes the quadratic Casimir invariant of $\Phi$. The pattern of
thermal inflation and relevant cosmological contents depend on in
what order the fields $X$, $S$ and $\ell$ roll down to the minimum.
Here we consider the case that the underlying theory
(\ref{UV-theory}) leads to \bea T_S \,<\, T_\ell \,<\, T_X, \eea
which is a natural possibility.

Meanwhile, the vanishing cosmological constant at the true vacuum
$\langle \varphi \rangle=\varphi_0$, $\langle \ell \rangle=0$
implies a nonzero potential energy near the origin: \bea V_0 &=&
\sum_{\varphi=X,S} V_\varphi = \sum_{\varphi=X,S} \alpha_\varphi^2
\hat m^2_\varphi \varphi^2_0, \eea with \bea \alpha_\varphi
&\approx& \frac{1}{\sqrt 2} \frac{m_{\sigma_\varphi}}{\hat m_\varphi},
\eea which has a value of ${\cal O}(10^{-1})$. Hence, during the
epoch of the thermal confinement, $V_0$ plays the role of vacuum
energy. If necessary to be definite, we fix the vacuum values of
the flaton fields as \bea X_0 \sim 10^{13}\,{\rm GeV}, &\quad& S_0
\sim 10^{12}\,{\rm GeV}, \eea where we have taken into account that
neutrino masses in the range of $10^{-2}$ eV are obtained
for $X_0 \sim \lambda_\nu\times 10^{15}$ GeV, and $\mu$ is
of ${\cal O}(m_{\rm soft})$ for $\hat\kappa S_0\sim X_0$.

\subsection{Thermal inflation}

When Hubble parameter $H$ becomes comparable to the mass of the
modulus $m_\phi$ at a time $t=t_\phi$, the modulus $\phi$ starts its
coherent oscillation with Planck scale amplitude, and the epoch of
moduli domination begins.

Thermal inflation begins at $t=t_1$ when the modulus energy density
$\rho_\phi$ becomes comparable to $V_0$, and thus one finds
\bea
\frac{\rho_\phi}{\rho_\mathrm{r}} &\sim& \frac{a(t_1)}{a(t_\phi)}
\sim \left(\frac{H(t_\phi)}{H(t_1)}\right)^{2/3}
\sim \left(\frac{m^2_\phi M^2_{Pl}}{V_0}\right)^{1/3},
\eea
where $\rho_r$ is the energy density of radiation, and the temperature
at $t_1$ is
\bea
T_1 &\sim& \left(\frac{\rho_\mathrm{r}}{\rho_\phi}\right)^{1/4}V^{1/4}_0
\sim \left(\frac{V^2_0}{m_\phi M_{Pl}}\right)^{1/6}.
\eea
This epoch of thermal inflation continues at least until $X$ becomes
unstable when the temperature drops to $T_X$ at $t=t_X$.

The thermal history after $t_X$ crucially depends on the ratio between $V_X$
and $V_S$.
If $V_X \gg V_S$, the Universe is dominated by the non-relativistic $X$
particles.
In this case, if $V_S$ becomes dominant before $S$ rolls down to its true
minimum, a second epoch of thermal inflation is driven by $V_S$ before or
after $\ell$ becomes unstable.
Otherwise, matter domination by $X$ particles would continue until the particles
decay and reheat the Universe for the Big-Bang nucleosynthesis.
On the other hand, if $V_X \lesssim V_S$, thermal inflation is driven essentially
by $V_S$, continuing until $S$ becomes unstable, and the Universe is reheated by
the decay of $S$.

\subsubsection{$V_X \gg V_S$}

Since the commencement of the oscillation of $X$, its energy density
$\rho_X$ dominates the Universe as matter.
As the Universe expands, $\rho_X$ becomes comparable to $V_S$ at $t=t_2$.

The total decay width of the radial flaton $\sigma_\varphi$ ($\varphi=X,S$)
can be written as
\bea
\Gamma_\varphi &=&  C_\varphi \frac{\hat m^3_\varphi}{\varphi^2_0},
\eea
where $C_\varphi$ has a value of ${\cal O}(10^{-3})$ because
$m^2_{\sigma_\varphi}={\cal O}(m^2_{\rm soft}/8\pi^2)$.
For $\ell$, one finds
\bea
\Gamma_\ell &=& C_\ell \frac{\hat m^3_\ell}{\ell^2_0},
\eea
with $C_\ell={\cal O}(10^{-1})$ and $\ell_0\sim \sqrt{|A_\nu X_0/\lambda_\nu|}$
being the vacuum value of $\ell$ in the absence of the $\mu$ term.
For $H \gg \Gamma_\varphi$, the contribution of $\rho_X$ to radiation at $t=t_2$
leads to\footnote{We assume that the energy loss of the oscillating field due
to the parametric resonance is small.}
\bea
T^4_{\rho_X}(t_2) &\simeq&
\left( \frac{\pi^2}{30} g_\ast(T_{\rho_X}(t_2)) \right)^{-1}
\left( \frac{\Gamma_X}{H(t_2)} V_S \right)
\sim
0.1V_S^{1/2} \Gamma_X M_{Pl}
\left( \frac{100}{g_\ast(T_{\rho_X})} \right),
\eea
which implies
\bea
\label{Teq-to-Tl}
\frac{T^4_{\rho_X}(t_2)}{T^4_\ell} &\sim&
10^{-6} \frac{M_{Pl}}{X_0}
\left( \frac{100}{g_\ast(T_{\rho_X}(t_2))} \right)
\left( \frac{C_X}{10^{-3}} \right)
\left( \frac{\alpha_S}{10^{-1}} \right)
\left( \frac{\hat m_S \hat m_X^3}{\hat m_\ell^4} \right)
\left( \frac{10 S_0}{X_0} \right).
\eea
For $X_0 \gtrsim 10^{12}$ GeV, $T_{\rho_X}(t_2)$ is lower than $T_\ell$
and thus the second thermal inflation can take place after $\ell$ becomes
unstable.
Meanwhile, in case of $T_{\rho_X}(t_2) < T_\ell$, the temperature due to
the radiation contribution of $\rho_\ell$ at $t=t_2$ is
\bea
\label{T2}
T^4_{\rho_\ell}(t_2)
&\simeq&
\left( \frac{\pi^2}{30} g_\ast(T_{\rho_\ell}(t_2)) \right)^{-1}
\left( \frac{\Gamma_\ell}{H(t_2)} \rho_\ell(t_2) \right)
\sim
0.1\frac{V_S^{1/2} V_\ell \Gamma_\ell \Gamma_X^2 M_{Pl}^3 }{T_\ell^8}
\left( \frac{100}{g_\ast(T_{\rho_\ell}(t_2))} \right),
\eea
and therefore, by using $V(\ell=0)-V(\ell=\ell_0)\sim \hat m^2_\ell\,\ell^2_0$
for $\mu=0$, we find
\bea
\label{T-to-TS}
\frac{T^4_{\rho_X}(t_2)}{T^4_{\rho_\ell}(t_2)} &\sim&
10^4
\frac{X^2_0}{M^2_{Pl}}
\left( \frac{10^{-1}}{C_\ell} \right)
\left( \frac{10^{-3}}{C_X} \right)
\left(\frac{\hat m_\ell}{\hat m_X} \right)^3,
\nonumber \\
\frac{T^4_{\rho_\ell}(t_2)}{T^4_S} &\sim&
10^{-10}\beta_S^4
\frac{M^3_{Pl}}{X^3_0}
\left( \frac{100}{g_\ast(t_2)} \right)
\left( \frac{\alpha_S}{10^{-1}} \right)
\left( \frac{C_\ell}{10^{-1}} \right)
\left( \frac{C_X}{10^{-3}} \right)^2
\left( \frac{\hat m_X^2}{\hat m_\ell \hat m_S} \right)^3
\left( \frac{10 S_0}{X_0} \right).
\eea
For $X_0 \lesssim 10^{15}$ GeV with $\beta_S \sim {\cal O}(1)$,
the partial decay of $\ell$ at $t=t_2$ thus contributes dominantly to
radiation with temperature larger than $T_S$.
This implies that, after $\ell$ decouples from thermal bath, a second epoch
of thermal inflation begins at $t=t_2$ with a background temperature
\bea
T_2 &=& T_{\rho_\ell}(t_2).
\eea
The thermal inflation ends as $S$ rolls away from the origin, and then
the decay of $S$ eventually reheats the Universe for the Big-Bang
nucleosynthesis\footnote{The contribution to radiation from the late decay
of $X$ is negligible for our choice of the flaton vacuum values.}

The number of $e$-folds for each epoch of thermal inflation is estimated as
\bea
N_1 &\simeq& \ln\left(\frac{T_1}{T_X}\right)
\simeq
6.8 + \ln\left[\beta_X^6
\left( \frac{\alpha_X}{10^{-1}} \right)^4
\left( \frac{X_0}{10^{13}{\rm GeV}} \right)^4
\left( \frac{10^3{\rm GeV}}{\hat m_X} \right)^2
\left( \frac{10^6{\rm GeV}}{m_\phi} \right) \right],
\nonumber \\
N_2 &\simeq&
\frac{1}{3}\ln\left(\frac{\rho_\ell(t_2)}{\rho_\ell(t_S)}\right)
\sim
\frac{1}{3}\ln\left[\frac{V_\ell}{T_\ell^4}
\frac{\Gamma_\ell M_{Pl}}{V_S^{1/2}}
\frac{V_S \Gamma_X^2 M_{Pl}^2}{T_\ell^8} \right]
\nonumber \\
&\simeq&
5.5 + \frac{1}{3}\ln\left[
\left( \frac{\alpha_S}{10^{-1}} \right)
\left( \frac{C_\ell}{10^{-1}} \right)
\left( \frac{C_X}{10^{-3}} \right)^2
\left( \frac{\hat m_S \hat m_X^6}{\hat m_\ell^7} \right)
\left( \frac{10 S_0}{X_0} \right)
\left( \frac{10^{13}{\rm GeV}}{X_0} \right)^3 \right].
\eea
These $e$-folds, which are additive to the standard primordial inflation,
do not interfere directly with CMB or cosmic 21{\it cm}
fluctuations\footnote{For primordial inflation at $V^{1/4} \sim 10^{10}$ GeV
and assuming radiation domination after inflation, if the reheating temperature
of thermal inflation is $T_{\rm d}={\cal O}(1)$ GeV, the observable Universe leaves
the horizon around 40 $e$-folds before the end of inflation.
The observed CMB covers only around 6 $e$-folds after the horizon exit of the
observable Universe \cite{Yao:2006px}, and the potentially observable cosmic
21{\it cm} fluctuations cover around 9 additional $e$-folds \cite{Kleban:2007jd}.},
but may have an observable effect on the primordial density perturbation though
it may be difficult to disentangle from uncertainties in the model of primordial
inflation.

The entropy productions from the decays of $\varphi$ and $\ell$ provide the
dilution factors
\bea
\Delta_X &\sim& \left( \frac{T_\ell}{T_X} \right)^3
\frac{V_X}{\rho_X(t_\ell)}
\sim \left( \frac{T_\ell}{T_X} \right)^3
\frac{V_X \Gamma_X^2 M_{Pl}^2}{T_\ell^8},
\\
\Delta_\ell
\label{ldfactor}
&\sim&
\left( \frac{T_S}{T_\ell} \right)^3 \frac{V_\ell}{\rho_\ell(t_S)}
\sim \left( \frac{T_S}{T_\ell} \right)^3
\frac{V_\ell \Gamma_\ell M_{Pl}}{T_S^4 V_S^{1/2}},
\\
\Delta_S
\label{Sdfactor}
&\sim&
\frac{V_S}{T_S^3 \,T_{\rm d}},
\eea
where $T_{\rm d}$ is the decay temperature of the flaton field $S$ after
thermal inflation.
Therefore, the total dilution is given by
\bea
\label{totaldfactor}
\Delta_{\rm tot}
&=& \Delta_X \, \Delta_\ell \, \Delta_S
\sim
\frac{V_X}{T_X^3 T_{\rm d}} \frac{V_\ell V_S^{1/2}
\Gamma_\ell \Gamma_X^2 M_{Pl}^3}{T_\ell^8 T_S^4}
\nonumber \\
&\sim&
10^{28} \,
\beta_X^3 \beta_S^4
\left( \frac{\alpha_X^2 \alpha_S}{10^{-3}} \right)
\left( \frac{C_\ell^2 C_X^4}{10^{-11} \, C_S} \right)^{1/2}
\left( \frac{10 S_0}{X_0} \right)^2
\left( \frac{\hat m_X^5}{\hat m_\ell^3 \hat m_S^2} \right)
\left( \frac{10^3 {\rm GeV}}{\hat m_S} \right)^{5/2}
\eea
where we have used $T_\mathrm{d} \sim \left(\Gamma_S M_{Pl} \right)^{1/2}$ though
we will redefine it later in a more definite way.

Thermal inflation should dilute enough the abundance of moduli to avoid dark matter
over-production described in Section \ref{Lsusy}.
The bound on the abundance at the time of the decay of moduli is
\bea
\label{moduliconstraint}
\frac{n_\phi}{s}
&\lesssim&
0.25 \frac{\Gamma_\phi}{\Gamma_{\phi \rightarrow \tilde{G} \tilde{G}}}
\frac{1}{m_\chi} \left( \frac{\rho_\mathrm{cr}}{s} \right)_\mathrm{present}
\simeq
5.0 \times 10^{-12} \frac{\Gamma_\phi}{\Gamma_{\phi \rightarrow
\tilde{G} \tilde{G}}} \left( \frac{100 \, \rm{GeV}}{m_\chi} \right).
\eea
For modulus particles produced before the thermal inflation, the late time
abundance is estimated as
\bea \label{bigbangmoduli}
\frac{n_\phi}{s} &\sim&
\left( \frac{M_{Pl}}{m_\phi} \right)^{1/2} \frac{1}{\Delta_{\rm tot}}
\sim
10^{-22} \left( \frac{10^6 {\rm GeV}}{m_\phi} \right)^{1/2}
\left( \frac{10^{28}}{\Delta_{\rm tot}} \right).
\eea
If produced at the end of the first thermal inflation, the modulus abundance is
given by
\bea \label{TI1moduli} \nonumber
\frac{n_\phi}{s}
&\sim&
\frac{m_\phi M_{Pl}^2}{T_X^3}
\frac{H(t_X)^4}{m_\phi^4} \frac{1}{\Delta_\mathrm{tot}}
\sim
\frac{V_X^2}{m_\phi^3 M_{Pl}^2 T_X^3}
\frac{1}{\Delta_{\rm tot}}
\\
&\sim& 10^{-32} \,
\beta_X^3
\left( \frac{\alpha_X}{10^{-1}} \right)^4
\left( \frac{\hat{m}_X}{10^3 \, {\rm GeV}} \right)
\left( \frac{10^6 \, {\rm GeV}}{m_\phi} \right)^3
\left( \frac{X_0}{10^{13} \, {\rm GeV}} \right)^4
\left( \frac{10^{28}}{\Delta_{\rm tot}} \right),
\eea
whereas, for those produced at the end of the second thermal inflation,
we find
\bea \label{TI2moduli} \nonumber
\frac{n_\phi}{s} &\sim&
\frac{m_\phi M_{Pl}^2}{T_S^3} \frac{H(t_S)^4}{m_\phi^4}
\frac{1}{\Delta_S}
\sim \frac{V_S T_{\rm d}}{m_\phi^3 M_{Pl}^2}
\\
&\sim& 10^{-26}
\left( \frac{\alpha_S}{10^{-1}} \right)
\left( \frac{C_S}{10^{-3}} \right)^{1/2}
\left( \frac{\hat{m}_S}{10^3 \, {\rm GeV}} \right)^{7/2}
\left( \frac{10^6 \, {\rm GeV}}{m_\phi} \right)^3
\left( \frac{S_0}{10^{12} \, {\rm GeV}} \right).
\eea
All of these contributions to the moduli abundance are far below the safe level
of (\ref{moduliconstraint}).

\subsubsection{$V_X \lesssim V_S$}

In this case, we have only single epoch of thermal inflation driven
essentially by $V_S$, but it is extended by the radiation contribution of $X$
and $\ell$ when those fields decouple from thermal bath.
The $e$-folding of the thermal inflation is given as
\bea
N &\simeq&
\ln\left[ \left( \frac{T_1}{T_X} \right)
\left( \frac{a_\ell}{a_X} \right)
\left( \frac{a_S}{a_\ell} \right)\right]
\sim
\frac{1}{3} \ln\left[
\frac{V_0}{m_\phi^{1/2} M_{Pl}^{1/2} T_X^3}
\frac{V_X \Gamma_X M_{Pl}}{T_\ell^4 V_S^{1/2}}
\frac{V_\ell \Gamma_\ell M_{Pl}}{T_S^4 V_S^{1/2}}
\right]
\nonumber \\
&\simeq&
12 + \frac{1}{3} \ln\left[
\beta_X^3 \beta_S^4
\left(\frac{\alpha_X}{10^{-1}}\right)^2
\left(\frac{C_\ell\, C_X}{10^{-4}}\right)
\left(\frac{\hat m_\ell \hat m_X^2}{\hat m_S^3}\right)
\left(\frac{10^3{\rm GeV}}{\hat m_S}\right)
\left(\frac{10^6{\rm GeV}}{m_\phi}\right)^{1/2} \right].
\eea
The dilution factor from the decay of $X$ is obtained as
\bea
\Delta_X &\sim&
\left( \frac{T_\ell}{T_X} \right)^3
\frac{V_X}{\rho_X(t_\ell)}
\sim
\left( \frac{T_\ell}{T_X} \right)^3 \frac{V_X}{T_\ell^4}
\frac{\Gamma_X M_{Pl}}{V_S^{1/2}},
\eea
while those from $\ell$ and $S$ have the same forms as
(\ref{ldfactor}) and (\ref{Sdfactor}), respectively.
Hence the total dilution factor is
\bea
\Delta_{\rm tot} &\sim&
\frac{V_X}{T_X^3 T_{\rm d}}
\frac{V_\ell \Gamma_\ell \Gamma_X M_{Pl}^2}{T_\ell^4 T_S^4}
\sim
\frac{\alpha_X^2 C_\ell\, C_X}{C_S^{1/2}} \beta_X^3 \beta_S^4
\left(\frac{\hat m_\ell \hat m_X^2}{\hat m_S^3}\right)
\left(\frac{S_0}{\hat m_S}\right)
\left(\frac{M_{Pl}}{\hat m_S}\right)^{3/2}
\nonumber \\
&\sim&
10^{28} \beta_X^3 \beta_S^4
\left( \frac{\alpha_X}{10^{-1}} \right)^2
\left( \frac{C_\ell^2 \, C_X^2}{10^{-5} \, C_S} \right)^{1/2}
\left(\frac{\hat m_\ell \hat m_X^2}{\hat m_S^3}\right)
\left(\frac{10^3{\rm GeV}}{\hat m_S}\right)^{5/2}
\left(\frac{S_0}{10^{12}{\rm GeV}}\right),
\eea
and thus, compared to the bound in (\ref{moduliconstraint}),
the abundance of moduli becomes totally negligible in this case too.

After thermal inflation, the Universe is reheated by the decay of the
flaton that ends the last thermal inflation.
Since there is no unique decay temperature, we instead define
the flaton decay temperature $T_{\rm d}$ by
\bea
\rho_\mathrm{r}(T_{\rm d}) &\equiv&
\frac{1}{2} \Gamma_\varphi^2 M_{Pl}^2,
\eea
or
\bea
\label{Td-def}
\frac{\pi^2}{30} g_\ast(T_{\rm d}) T_{\rm d}^4 &=&
\rho_{\rm SM}(T_{\rm d})
= \frac{1}{2} \Gamma_{\sigma_\varphi \to{\rm SM}}
\Gamma_\varphi M_{Pl}^2,
\eea
where $\Gamma_{\sigma_\varphi\to{\rm SM}}$ is the decay width to
the SM particles, and $\varphi$ is either $X$ or $S$.
This corresponds to a time
\bea
t_{\rm d} &\simeq& \frac{1}{\Gamma_\varphi},
\eea
and at the moment we have
\bea
\rho_\varphi(T_{\rm d}) &\simeq&
\frac{1}{2} \Gamma_\varphi^2,
\eea
and the entropy increases after $t_{\rm d}$ by a factor
$S_{\rm f}/S_{\rm d}\simeq 2$ where $S_{\rm d}$ and $S_{\rm f}$ are
the entropy at $t=t_{\rm d}$ and a late time, respectively \cite{Kim:2008yu}.

\subsection{Baryogenesis}

For $T_X>T_\ell>T_S$, the flaton field $X$ decouples first from thermal
bath and settles to its true minimum $X=X_0$.
The scalar potential implementing the AD leptogenesis is then determined
by the effective theory (\ref{Model}), which involves
\bea
\label{Model-AD}
{\cal L}_{\rm AD} &=&
\int d^4\theta\, Y^{\rm eff}_S S S^*
+ \left(
\int d^4\theta\,\hat \kappa \frac{S^*}{X_0} H_u H_d
+ \int d^2\theta\,
\frac{1}{2}\lambda_\nu\frac{LH_uLH_u}{X_0} + {\rm h.c.} \right),
\eea
where $Y^{\rm eff}_S=Y_S(Q=|S|)$.
To analyze the dynamics induced by $m^2_{\tilde L}(Q) + m^2_{H_u}(Q)<0$ at
$Q\sim m_{\rm soft}$, we parameterize the associated flat
directions\footnote{We set quark and lepton directions to be zero since they
are expected to be held at the origin throughout the dynamics.
This will become clear from the subsequent argument.} as
\bea
L = (0,l)^T, \qquad H_u = (h_u,0)^T, \qquad H_d = (0,h_d)^T
\eea
with the $D$-term constraint for the $SU(2)_L$ gauge symmetry
\bea
D_2 &=& |h_u|^2 - |h_d|^2 - |l|^2 = 0.
\eea
The relevant part of the potential is then written as
\bea
V_{\rm AD} &=& V_S + \frac{1}{2}g^2_2 D^2_2
+ m^2_S |S|^2 + m^2_{\tilde L} |l|^2
+ m^2_{H_u}|h_u|^2 + m^2_{H_d}|h_d|^2
\nonumber \\
&&
+ \left( \frac{1}{2}A_\nu \lambda_\nu l^2 h^2_u
- B\hat\mu \frac{S^*}{X_0}h_uh_d + {\rm c.c.}\right)
\nonumber \\
&&
+ \left|\lambda_\nu l h^2_u \right|^2
+ \left|\lambda_\nu  l^2 h_u + \hat\mu \frac{S^*}{X_0}h_d \right|^2
+ \left|\,\hat\mu \frac{S^*}{X_0}h_u\right|^2,
\eea
with $\hat\mu$ given by
\bea
\hat\mu &=& \hat\kappa\left(
\left(\frac{F^S}{S}\right)^*
+ {\cal O}(m_{\rm soft}) \right),
\eea
where the loop-suppressed contribution from $F^C$ is included
in ${\cal O}(m_{\rm soft})$.
Note that $B$, $A_\nu$, and $\hat\mu$ have values of ${\cal O}(m_{\rm soft})$
and are nearly independent of $\varphi=X,S$ for $|\varphi|\gg m_{\rm soft}$
since the equation of motion for $F^\varphi$ is given by (\ref{F-EOM}).

Initially, all fields are held at the origin by their finite temperature
potential.
As the temperature drops down, one of the unstable directions, $S$ or $\ell=lh_u$,
will roll away from the origin.
We assume $lh_u$ rolls away first, i.e. $T_\ell<T_S$.
Then the potential term $\frac{1}{2}A_\nu\lambda_\nu l^2 h_u^2$ fixes the phase of
$lh_u$, while $|\lambda_\nu l h_u^2|^2$ and $|\lambda_\nu l^2 h_u|^2$ stabilize
its magnitude.
The $lh_u$ field may partially reheat the thermal bath and so prolong the thermal
inflation, but eventually $S$ will also roll away from the origin, ending
thermal inflation.
As $S$ rolls away, the term $B\hat\mu S^*h_u h_d/X_0$ will force $h_d$
to become non-zero.
This provides temporarily a large mass to quark and lepton directions and
constrains those directions to zero, shielding the dynamics from the dangerous
non-MSSM vacuum in the direction associated quark and lepton
\cite{Kim:2008yu,Casas:1995pd}.
Then, $B\hat\mu S^*h_u h_d/X_0$ fixes the phase of $S^* h_u h_d$.
As $S$ nears its minimum, the cross term from
$|\lambda_\nu l^2h_u + \hat\mu S^* h_d/X_0|^2$ rotates the phase of $lh_u$
generating a lepton asymmetry, and at the same time $|\hat\mu S^* h_u/X_0|^2$
gives an extra contribution to the mass squares of $lh_u$ and $h_uh_d$,
bringing them back in towards the origin.
Thus, we have a type of the Affleck-Dine (AD) leptogenesis.
Preheating then damps the amplitude of the $lh_u$ and $h_uh_d$ fields keeping them
in the lepton preserving region near the origin \cite{Felder:2007iz}.
The $lh_u$ and $h_uh_d$ fields then decay, at a temperature in the MSSM sector
above the electroweak scale, and their lepton number is converted to baryon number
by sphaleron processes.
Finally, the flatons $S$ and $X$ decay, diluting the baryon density to the value
required by observations, $n_B/s \sim 10^{-10}$.

One may think that, if $X$ decays later than $S$, the baryon asymmetry generated
due to the dynamics of $S$ may be significantly diluted.
However, the dilution is possible only when $V_X \ggg V_S$ with $X_0$ different
from $S_0$ by more than several order,
which is very unnatural in our model.

Comparing to the model considered in \cite{Kim:2008yu}, the Higgs $\mu$ term
in our model has a linear dependence on the triggering field $S$ rather
than quadratic:
\beq
\label{AD-mu}
\mu = \hat \mu \frac{S^*}{X_0},
\eeq
where $\hat\mu$ does not depend on $S$ for $|S|\gg m_{\rm soft}$.
This may weaken the strength of torque responsible for the angular
momentum and preheating, but would only make a difference of a factor of
${\cal O}(1)$.
We therefore expect that the AD baryogenesis would work well in our model too.

\subsection{Dark matter}

The model (\ref{Model}) provides the lightest flatino as the dark
matter under the assumption of R-parity conservation since flatinos
$\psi_\varphi$ ($\varphi=X,S$) have small masses of
${\cal O}(m_{\rm soft}/8\pi^2)$.
In addition, the axions $a_\varphi$ can also become a dark matter of the
Universe if they are light enough to be stable.
These dark matter components can be cold, warm or even hot, depending on
their masses and how they are produced.
In this subsection, we will derive cosmological constraints on the dark
matter.

\subsubsection{Cold dark matter}

As a cold dark matter, the lightest flatino is dominantly produced
by the decay of the corresponding flaton if it eventually reheats
the Universe for the Big-Bang nucleosynthesis.
Another source is the decay of heavier sparticles in thermal bath,
which is dominated by the decay of the lightest MSSM sparticle.
In addition, axion misalignment and strings also contribute to the
energy density of cold dark matter.

\paragraph{Flatinos produced by the flaton decay:}
\label{flatonflatinos}

For the flaton $\varphi$ that ends the thermal inflation, its
decay reheats the Universe.
The late time flatino abundance from the flaton decay is determined
by \cite{Kim:2008yu}
\bea
\label{flatonaxinoyield}
\frac{n_{\psi_\varphi}}{s} &=&
\frac{2 \Gamma_{\sigma_\varphi \to \psi_\varphi \psi_\varphi}}
{m_{\sigma_\varphi} a^3}
\int_0^t d t^\prime\, a^3(t^\prime)\rho_\varphi(t^\prime)
\simeq
\frac{2.2 T_{\rm d}}{m_{\sigma_\varphi} }
\frac{\Gamma_{\sigma_\varphi\to \psi_\varphi\psi_\varphi}}
{\Gamma_{\sigma_\varphi\to{\rm SM}}},
\eea
where the decay temperature $T_{\rm d}$ is roughly given by
\bea
T_{\rm d} &\sim& 1{\rm GeV}
\left[
\left(\frac{100}{g_\ast(T_{\rm d})}\right)
\left(\frac{C_\varphi}{10^{-3}}\right)^2
\left(\frac{\hat m_\varphi}{10^3{\rm GeV}}\right)^6
\left(\frac{10^{12}{\rm GeV}}{\varphi_0}\right)^4
\right]^{1/4},
\eea
which follows from (\ref{Td-def}).
Using (\ref{Flaton-decay-rate}) and (\ref{flatonaxinoyield}),
one can find the current abundance of flatino dark matter $\psi_0$
\bea
\Omega_{\psi_0}
&\simeq&
5.6 \times 10^8 \left( \frac{m_{\psi_0}}{1{\rm GeV}} \right)
\frac{n_{\psi_\varphi}}{s}
\nonumber \\
&\simeq&
3.6 \left(\frac{100}{g_\ast(T_{\rm d})}\right)^{1/2}
\left( \frac{\lambda_{\psi_\varphi}}{10^{-1}} \right)^2
\left( \frac{m_{\psi_0}}{1{\rm GeV}} \right)
\left(\frac{m_{\psi_\varphi}}{1{\rm GeV}}\right)^2
\left( \frac{1{\rm GeV}}{T_{\rm d}} \right)
\left( \frac{10^{11}{\rm GeV}}{\varphi_0} \right)^2,
\eea
where we have used that one flatino dark matter is produced per each heavier
flatino $\psi_1$ if $\psi_\varphi=\psi_1$, and that the total decay width
is given by $\Gamma_\varphi\simeq \Gamma_{\sigma_\varphi\to{\rm SM}}$.
Therefore, the requirement $\Omega_{\psi_0}\leq \Omega_{\rm CDM}\simeq 0.2$
translates into
\bea
\label{maxinoflatoncon}
m_{\psi_0} &\lesssim&
1.8{\rm GeV}
\left[
\left( \frac{g_\ast({T_{\rm d}})}{100} \right)^2
\left( \frac{T_{\rm d}}{1 {\rm GeV}} \right)
\left( \frac{10^{-1}}{\lambda_{\psi_\varphi}} \right)^2
\left( \frac{\varphi_0}{10^{12}{\rm GeV}} \right)^2
\right]^\frac{1}{3},
\eea
which is at the lower end of its expected range (\ref{SX-masses}).

\paragraph{Flatinos produced by the decay of thermally generated MSSM sparticles:}

The flaton coupling to $H_uH_d$ in the effective K\"ahler potential
(\ref{Model}) induces
\bea
{\cal L}_{\rm int} &=&
\lambda_\kappa \frac{\mu}{S_0} H_{u,d}\tilde H_{d,u}\psi_S
+ i\frac{\lambda^\prime_\kappa}{S_0}H_{u,d}\tilde H_{d,u}\sigma^\mu
\partial_\mu \bar \psi_S
- \frac{\mu}{X_0}H_{u,d}\bar H_{d,u}\psi_X
+ {\rm h.c},
\eea
where $\lambda_\kappa=\langle S\partial_S \ln \hat\kappa \rangle
={\cal O}(1/8\pi^2)$,
$\lambda^\prime_\kappa=\hat\kappa S_0/X_0={\cal O}(1)$, and
$\tilde H_{u,d}$ are the Higgsino fields.
The above interactions lead the lightest MSSM sparticle $\chi$ to
decay into flatinos with the decay rate \cite{Covi:1999ty}
\bea
\label{gammahiggsino}
\Gamma_{\chi \to \psi_\varphi} &=& \frac{C_{\psi_\varphi}}{16\pi}
\frac{m^3_\chi}{\varphi_0^2},
\eea
where $C_{\psi_S}={\cal O}(10^{-2}C_{\psi_X})$ for $\mu$ of the soft mass
scale and $S_0\sim 10^{-1}X_0$.
Here $C_{\psi_X}\sim 1$ may contain a factor of $m^2_Z/m^2_\chi$, and
we have neglected the masses of decay products.

The thermal bath generates $\chi$ with the number density
\bea
n_\chi &=& \frac{1}{\pi^2}
\int_0^\infty d k \frac{k^2}{\exp\sqrt{\frac{k^2+m_\chi^2}{T^2}} + 1},
\eea
and they subsequently decay into flatinos, dominantly into $\psi_X$.
The late time flatino abundance is thus estimated as
\bea
\label{ndecay}
\frac{n_{\psi_X}}{s} &=&
\frac{S_{\rm d}}{S_{\rm f}} \frac{1}{s_{\rm d}}
\int_0^\infty d t \left( \frac{a(t)}{a_{\rm d}} \right)^3
n_\chi(t) \Gamma_{\chi \to \psi_X},
\eea
whose numerical solution can be found in \cite{Kim:2008yu}
\bea
\label{naxino}
\frac{n_{\psi_X}}{s} &=&
g^{-1/4}_\ast(T_{\rm d}) g^{-5/4}_\ast(T_\chi)
\left(\frac{\Gamma_{\sigma_\varphi\to{\rm SM}}}{\Gamma_\varphi}\right)^{1/2}
\frac{\Gamma_{\chi\to\psi_X} M_{Pl}}{m^2_\chi} F_\psi(x),
\eea
where $F_\psi(x)$ is approximated as
\bea
F_\psi(x) &\sim&
\left\{
\begin{array}{ccc}
5.3 x^7 & {\rm for} & x \ll 1
\\
5.4 & {\rm for} & x \gg 1
\end{array}
\right.,
\eea
with
\bea
x &=& \frac{2}{3}
\left(\frac{g_\ast(T_{\rm d})}{g_\ast(T_\chi)}\right)^{1/4}
\frac{T_{\rm d}}{T_\chi},
\eea
and $T_\chi\simeq 2m_\chi/21$ being the temperature at which the flatino
production rate is maximized.
Using (\ref{gammahiggsino}) and (\ref{naxino}), the current abundance of
flatino dark matter is obtained
\bea
\label{omegaaxinoflatoncon}
\Omega_{\psi_0} &\simeq& 5.6\times 10^8
\left(\frac{m_{\psi_0}}{1{\rm GeV}}\right) \frac{n_{\psi_X}}{s}
\nonumber \\
&\simeq&
2.7\times 10^{-2} C_{\psi_X} \Big(
\frac{10^3}{g^{1/4}_\ast(T_{\rm d})g^{5/4}_\ast(T_\chi)}\Big)
\left(\frac{m_\chi}{10^2{\rm GeV}}\right)
\left(\frac{m_{\psi_0}}{1{\rm GeV}}\right)
\left(\frac{10^{13}{\rm GeV}}{X_0}\right)^2
F_\psi(x),
\eea
where we have used $\Gamma_\varphi\simeq \Gamma_{\sigma_\varphi\to{\rm SM}}$,
and that the number density of $\psi_0$ is the same as
that of $\psi_X$ even when $\psi_X=\psi_1$.
Therefore, for $x\ll 1$, one finds
\bea
\label{approxomeganlsp}
\Omega_{\psi_0} \simeq 1.2\times 10^5\, C_{\psi_X}
\left(\frac{T_{\rm d}}{m_\chi}\right)^7
\Big(\frac{10^3 g^{3/2}_\ast(T_{\rm d})}{g^3_\ast(T_\chi)}\Big)
\left(\frac{m_\chi}{10^2{\rm GeV}}\right)
\left(\frac{m_{\psi_0}}{1{\rm GeV}}\right)
\left(\frac{10^{13}{\rm GeV}}{X_0}\right)^2,
\eea
which does not exceed $\Omega_{\rm CDM}$ if the flaton decay temperature
satisfies
\bea
T_{\rm d} \lesssim \frac{m_\chi}{6.7}
\left(\frac{g_\ast(T_\chi)}{g_\ast(T_{\rm d})}\right)^{1/4}
\left[C^{-1}_{\psi_X}
\Big(\frac{g^{1/4}_\ast(T_{\rm d})g^{5/4}_\ast}{10^3}\Big)
\left(\frac{10^2{\rm GeV}}{m_\chi}\right)
\left(\frac{1{\rm GeV}}{m_{\psi_0}}\right)
\left(\frac{X_0}{10^{13}{\rm GeV}}\right)^2 \right]^{1/7}.
\eea
Since $T_{\rm d}$ is expected to be a few GeV, the above cosmological bound is
well satisfied for $m_\chi={\cal O}(10^2)$ GeV.
Note that the flaton decay temperature can be less than or similar to the
freeze-out temperature of $\chi$, which is about $m_\chi/20$
\cite{Yao:2006px}.
Therefore, in order to avoid direct production of $\chi$ from the flaton
decay, we may require $m_{\sigma_\varphi}< 2m_\chi$, which is quite plausible
in our model.

Flatinos will also be produced by the decay of $\chi$ after they freeze out.
However, the standard Big-Bang neutralino freeze-out abundance is good match
to the dark matter abundance, our freeze-out abundance of $\chi$ will
typically be less than the standard abundance, and $m_{\psi_\varphi}\ll m_\chi$,
therefore the flatino abundance generated after the freeze-out should be safe.

\paragraph{Axion misalignment:}

The cold dark matter contributions of misalignment and axionic strings
from QCD-axion are well known.
Thus we consider only the cases of non-QCD axion under the assumption that
the mass of the axion appears due to a tree-level symmetry breaking term
which is small enough not to disturb the radiative stabilization of the
associated flaton field.
In this case, the mass of axion is not connected with its coupling constant.

The energy density of an axion misalignment when oscillation commences is
\bea
\rho_a &=&  \frac{m_a^2 \Theta^2}{2} \frac{\varphi^2_0}{{\cal N}^2},
\eea
where $a=a_\varphi$, $m_a$ is the axion mass, $\Theta$ is the misalignment
angle, and ${\cal N}$ is the vacuum degeneracy.
The current energy density is
\bea
\frac{\rho_a}{\rho_{\rm r}}
&=&
\left( \frac{a_0}{a_{\rm osc}} \right)
\frac{g_\ast(T_{\rm osc})}{g_\ast(T_0)}
\left( \frac{g_{\ast s}(T_0)}{g_{\ast s}(T_{\rm osc})} \right)^{4/3}
\left( \frac{\rho_a}{\rho_{\rm r}} \right)_{\rm osc}
\\
&=&
\frac{\sqrt{3}}{2}
\left( \frac{\pi^2}{9} \right)^{1/4}
\frac{g^{5/4}_\ast(T_{\rm osc})}{g_\ast(T_0)}
\left( \frac{g_{\ast s}(T_0)}{g_{\ast s}(T_{\rm osc})} \right)^{1/3}
\frac{\Theta^2}{{\cal N}^2}
\frac{m_a^2 \varphi^2}{\left(3 H_{\rm osc} M_{Pl}\right)^{3/2} T_0}.
\eea
Therefore, $\Omega_a < \Omega_{\rm CDM}$ requires
\bea
\label{mamisbound}
m_a &<& 6.0 \times 10^{-5} {\rm eV}
\left( \frac{100}{g_\ast(T_{\rm osc})}\right)^{11/6}
\frac{{\cal N}^4}{\Theta^4}
\left( \frac{10^{12}{\rm GeV}}{\varphi_0} \right)^4,
\eea
where we have used $3H_{\rm osc} = m_a$ and
$g_\ast(T_{\rm osc})=g_{\ast s}(T_{\rm osc})$.

Thermal inflation can dilute those contributions if the reheating temperature
is lower than the temperature at which the mass of axion becomes comparable
to the expansion rate.
We do not analyze the dilution in this paper, instead we refer the reader
to \cite{Kim:2008yu} for the case.

\paragraph{Axionic strings:}

The axion produced by the strings is \cite{KT}
\bea
\frac{n_a}{s} &\sim& A \frac{\varphi_0^2}{T_{\rm osc} M_{Pl}},
\eea
where $A \equiv \left[1\, {\rm or }\,
\ln{\left(\hat \mu/H_{\rm osc}\right)} \right]$.
The current energy density is
\bea
\frac{\rho_a}{\rho_{\rm r}}
&=&
\frac{4}{3}
A \left( \frac{\pi^2}{10} g_\ast(T_{\rm osc}) \right)^{1/4}
\left( \frac{g_{\ast s}(T_0)}{g_\ast(T_0)} \right)
\frac{m_a \varphi_0^2}{3 H_{\rm osc}^{1/2} M_{Pl}^{3/2} T_0}.
\eea
Therefore, in order for $\Omega_a$ not to exceed $\Omega_{\rm CDM}$,
it is required
\bea
\label{mastringbound}
m_a &\lesssim&
2.5 \times 10^{-6} {\rm eV}
\left( \frac{50}{A} \right)^2
\left( \frac{100}{g_\ast(T_{\rm osc})}\right)^{1/2}
\left( \frac{10^{12}{\rm GeV}}{\varphi_0} \right)^4,
\eea
where we have again used $3H_{\rm osc} = m_a$.

The mass bound (\ref{mastringbound}) implies that tree level symmetry
breaking should be very small.
For instance, the higher dimensional superpotential term
\bea
\Delta W &\propto& \frac{\varphi^n}{M^{n-3}_{Pl}}, \quad(n>3)
\eea
provides the axion a mass
\bea
\label{PNGBmass}
m_a^2 &\sim&
\left| A \varphi_0 \right|
\left(\frac{\varphi_0}{M_{Pl}}\right)^{n-3},
\eea
where $A \sim m_{3/2}$ (or $m_{\rm soft}$).
In this case, the requirement (\ref{mastringbound}) is satisfied for
\bea
n &\gtrsim& 3 +
\frac{\ln(m^2_a/|A\varphi_0|)}{\ln(\varphi_0/M_{Pl})}
\nonumber \\
&\gtrsim&
7.4
- \frac{1}{15} \ln\left[\left(\frac{m_a}{10^{-6}{\rm eV}}\right)^2
\left( \frac{10^4 {\rm GeV}}{|A|} \right)
\left(\frac{10^{12}{\rm GeV}}{\varphi_0}\right) \right]
- \frac{1}{3}
\ln\left(\frac{\varphi_0}{10^{12}{\rm GeV}}\right).
\eea

\subsubsection{Warm/hot dark matter}

The LSP $\psi_0$ and axion can be warm or hot when they are produced by
the decay of the next to LSP (NLSP) $\psi_1$ (for LSP and axion) and the
radial flaton $\sigma_\varphi$ (for axion).
The constraint on hot dark matter comes from CMBR and structure formation
\cite{Bashinsky:2003tk,Hannestad:2003ye,Crotty:2004gm,Hannestad:2008js}.
Currently allowed hot dark matter fractional contribution to the present
critical density is $\Omega_{HDM} \lesssim 10^{-2}$ \cite{Amsler:2008zzb}.
The constraint may be more stringent as suggested from the analysis of
the early re-ionization of the Universe at high redshift \cite{Jedamzik:2005sx}.
Taking into account the recent analysis of WMAP $5$-year data
\cite{Dunkley:2008ie}, the allowed warm/hot dark matter fractional
contribution to the present critical density is likely to be
\bea
\label{WHDMconstraint}
\Omega_{\rm WHDM} &\lesssim& 10^{-3}.
\eea
We will take this as the upper bound on the fractional energy density of
our warm/hot dark matter.

\paragraph{Hot axion from the flaton decay:}

Axions produced by the flaton decay have a current momentum
\bea
\label{pa}
p_a &=& \frac{a}{a_0}\frac{m_{\sigma_\varphi}}{2},
\eea
where $a$ is the scale factor at the time they were created and $a_0$
is the scale factor now.
Whereas, the current momentum of an axion produced at $t_{\rm d}$ is
\bea
p_{\rm d} &=& \frac{a_{\rm d}}{a_0} \frac{m_{\sigma_\varphi}}{2}
= \frac{S_{\rm d}^{1/3} g_{\ast S}^{1/3}(T_0)T_0}
{S_{\rm f}^{1/3} g_{\ast S}^{1/3}(T_{\rm d})T_{\rm d}}
\frac{m_{\sigma_\varphi}}{2}
\simeq 1.48 \times 10^{-4}{\rm eV} \Big(
\frac{m_{\sigma_\varphi}}{g_\ast^{1/3}(T_{\rm d})T_{\rm d}} \Big),
\eea
so it may be relativistic now.
The current number density spectrum is given by
\bea
p_a \frac{dn_a^{\rm hot}}{dp_a} &=&
\left( \frac{a}{a_0} \right)^3
\frac{2 \rho_\varphi}{m_{\sigma_\varphi}}
\frac{\Gamma_{\sigma_\varphi \to a a}}{H}
= \frac{16  p_{\rm d}^3}{m^4_{\sigma_\varphi}}
\frac{\Gamma_{\sigma_\varphi \to a a} p_a^3}{H p_{\rm d}^3}\rho_\varphi,
\eea
which may provide an observational test of our model in the future.
The energy density of the axions is
\bea
\label{rhoaflaton}
\frac{\rho_a^{\rm hot}}{\rho_{\rm SM}}
&=&
\frac{g_\ast(T_{\rm d}) g_{\ast S}^{4/3}(T)}
{g_\ast(T) g^{4/3}_{\ast S}(T_{\rm d})}
\frac{\Gamma_{\sigma_\varphi\to a a}}{\Gamma_{\sigma_\varphi\to {\rm SM}}}.
\eea
Therefore, assuming that the hot axions are still relativistic now,
their current energy density is estimated as\footnote{The energy density of
thermally produced axions is $\Omega_a \sim m_a/131{\rm eV}$,
hence it will be subdominant \cite{KT}.}
\bea
\label{omegaaflaton}
\Omega_a^\mathrm{hot} &\simeq& 2 \times 10^{-5}
\left( \frac{100}{g_\ast(T_{\rm d})} \right)^\frac{1}{3}
\frac{\Gamma_{\sigma_\varphi\to a a}}{\Gamma_{\sigma_\varphi\to {\rm SM}}}.
\eea

\paragraph{Hot axion from the decay of the NLSP:}

Because of the coupling (\ref{Singlino-interaction}), axions can be produced
also by the decay of the NLSP, i.e. the heavier flatino $\psi_1$, which is
originated from the decays of flaton and $\chi$.
Neglecting axion mass, the current momentum of the axion from the
NLSP\footnote{We ignore the effect of NLSP's momentum, since NLSPs from the
decay of flatons and $\chi$ will be highly non-relativistic during most of
its life.}
is given by
\bea
\label{NLSPacmomentum-t}
p_a &=& \frac{a}{a_0} \frac{m_{\psi_1}^2 - m_{\psi_0}^2}{2 m_{\psi_1}}.
\eea
In particular, the current momentum of the axion produced at
$t^\prime_{\rm d} = \Gamma_{\psi_1\to\psi_0}^{-1}$ is
\bea
\label{NLSPacmomentum-decay}
p^\prime_{\rm d} &=&
\frac{a^\prime_{\rm d}}{a_0}
\frac{m_{\psi_1}^2 - m_{\psi_0}^2}{2m_{\psi_1}}
=
\frac{g_{\ast S}^{1/3}(T_0) T_0}
{g_{\ast S}^{1/3}(T^\prime_{\rm d})T^\prime_{\rm d}}
\frac{m_{\psi_1}^2 - m_{\psi_0}^2}{2 m_{\psi_1}}
\nonumber \\
&\simeq& 1.87 \times 10^{-4} {\rm eV}
\left( \frac{m_{\psi_1}}{g^{1/3}_\ast(T^\prime_{\rm d})T^\prime_{\rm d}}
\right)
\left( 1 - \frac{m_{\psi_0}^2}{m_{\psi_1}^2} \right),
\eea
where $T^\prime_{\rm d}$ is the background temperature when the NLSP
decays
\bea
\label{TNLSPdecay}
T^\prime_{\rm d} &\sim&
\left( \frac{\pi^2}{90} g_\ast(T^\prime_{\rm d}) \right)^{-1/4}
\left( \Gamma_{\psi_1\to\psi_0} M_{Pl} \right)^{1/2}
\sim
10 {\rm eV}
\left(\frac{\Gamma_{\psi_1\to\psi_0}}{10^{-35}{\rm GeV}}\right)^{1/2}.
\eea
In this case, the axion would thus be ultra-relativistic now for any
plausible axion mass.
The current number density spectrum is
\bea
\label{NLSPanspectrum}
p_a \frac{dn_a^{\rm hot}}{dp_a} &=&
\left( \frac{a}{a_0} \right)^3 \frac{2 \rho_{\psi_1}}{m_{\psi_1}}
\frac{\Gamma_{\psi_1\to\psi_0}}{H}
= \frac{16 {p^\prime_{\rm d}}^3}{m_{\psi_1}^4}
\rho_{\psi_1} \frac{\Gamma_{\psi_1\to\psi_0} p_a^3}{H {p^\prime_{\rm d}}^3},
\eea
which again may provide an observational test of our model in the future.
The late time energy densities of the axions are
\bea
\label{rhoaflatonNLSP}
\left. \frac{\rho_a^{\rm hot}}{\rho_{\rm SM}} \right|_{\rm flaton}
&\simeq&
\left(\frac{g_\ast(T_{\rm d}) g^{4/3}_{\ast S}(T)}
{g_\ast(T)g_{\ast S}(T_{\rm d})g^{1/3}_{\ast S}(T^\prime_{\rm d})} \right)
\frac{T_{\rm d}}{T^\prime_{\rm d}}
\frac{m_{\psi_1}}{m_{\sigma_\varphi}}
\frac{\Gamma_{\varphi\to \psi_1\psi_1}}{\Gamma_{\sigma_S\to{\rm SM}}},
\\ \label{rhoaLOSPNLSP}
\left.\frac{\rho^{\rm hot}_a}{\rho_{\rm SM}}\right|_{\chi} &\simeq&
\left(\frac{g^{4/3}_{\ast S}(T)}{g_\ast(T)g^{1/3}_{\ast S}(T^\prime_{\rm d})}
\right) \frac{m_{\psi_1}}{T^\prime_d}
\left.\left(\frac{n_{\psi_1}}{s}\right)\right|_{t=t^\prime_d},
\eea
where $(n_{\psi_1}/s)_{t^\prime_{\rm d}}$ is given by (\ref{naxino}),
and their current energy densities are
\bea
\label{omegaaflatonNLSP}
\left. \Omega_a^{\rm hot} \right|_{\rm flaton}
&\simeq&
4 \times 10^{-5} \left(\frac{10}{g_\ast(T^\prime_{\rm d})}\right)^\frac{1}{3}
\frac{T_{\rm d}}{T^\prime_{\rm d}}
\frac{m_{\psi_1}}{m_{\sigma_\varphi}}
\frac{\Gamma_{\varphi\to \psi_1\psi_1}}{\Gamma_{\sigma_S\to{\rm SM}}}
\\ \label{omegaaLOSPNLSP}
\left. \Omega_a^{\rm hot} \right|_\chi
&\simeq&
1.8 \times 10^{-4}
\left( \frac{10}{g_\ast(T^\prime_{\rm d})} \right)^\frac{1}{3}
\left( \frac{10^3}{g^{1/4}_\ast(T_{\rm d}) g^{5/4}_\ast(T_\chi)}\right)
\frac{m_{\psi_1}}{T^\prime_{\rm d}}
\frac{\Gamma_{\chi\to\psi_1}M_{Pl}}{m^2_\chi}
\left(\frac{T_{\rm d}}{T_\chi}\right)^7.
\eea
The energy densities in (\ref{omegaaflaton}), (\ref{omegaaflatonNLSP}) and
(\ref{omegaaLOSPNLSP}) are typically well below the bound of
(\ref{WHDMconstraint}) in our model.

\paragraph{Warm/hot LSP from the decay of NLSP:}

Eqs.~(\ref{NLSPacmomentum-t}) - (\ref{NLSPanspectrum}) are applicable
in this case too.
Note that, although it is heavy and non-relativistic now, the LSP may
be warm or even hot unless the NLSP ($\psi_1$) and LSP ($\psi_0$) are
highly degenerate, and may have observable astrophysical effects.
The late time energy density of the LSP is simply given by
\bea
\Omega_{\psi_0} &=& \frac{m_{\psi_0}}{m_{\psi_1}} \Omega_{\psi_1},
\eea
where $\Omega_{\psi_1}$ is the would-be energy density of the NLSP $\psi_1$ if
it did not decay.
Therefore, from (\ref{maxinoflatoncon}), (\ref{omegaaxinoflatoncon}),
(\ref{approxomeganlsp}), (\ref{WHDMconstraint}) and (\ref{anaxioncon}),
$\Omega_{\psi_0} \lesssim \Omega_\mathrm{WHDM}$ requires for NLSPs from
flaton
\bea
m_{\psi_1} \lesssim 0.3 {\rm GeV} \left[
\left( \frac{m_{\psi_1}}{m_{\psi_0}} \right)
\frac{\Gamma_{\sigma_\varphi\to {\rm SM}}^{1/2}}{\Gamma_\varphi^{1/2}}
\left( \frac{g_\ast^{1/2}(T_{\rm d})}{10} \right)
\left( \frac{T_{\rm d}}{1 {\rm GeV}} \right)
\left( \frac{10^{-1}}{\lambda_{\psi}} \right)^2
\left( \frac{\varphi_0}{10^{12} {\rm GeV}} \right)^2
\right]^\frac{1}{3},
\eea
while for those from $\chi$
\bea
\label{anaxioncon}
T_{\rm d} \lesssim \frac{m_\chi}{26}
\left( \frac{g_\ast(T_\chi)}{g_\ast(T_{\rm d})} \right)^\frac{1}{4}
\left[\frac{1}{A}
\frac{g_\ast^{1/4}(T_{\rm d}) g_\ast^{5/4}(T_\chi)}{10^3}
\left( \frac{10^2 {\rm GeV}}{m_\chi} \right)
\left( \frac{1{\rm GeV}}{m_{\psi_0}} \right)
\left( \frac{\varphi_0}{10^{12}{\rm GeV}} \right)^2
\right]^\frac{1}{7},
\eea
where $\Gamma_\varphi\simeq \Gamma_{\sigma_\varphi \to {\rm SM}}$ has
been used.

\section{Conclusion}
Heavy gravitino with $m_{3/2}={\cal O}(10)$ TeV is a generic
prediction of sequestered SUSY breaking scenario. String flux
compactification provides a natural setup for sequestered SUSY
breaking, i.e. SUSY breaking at the IR end of warped throat, and
also can stabilize all moduli by fluxes or nonperturbative effects.
The resulting moduli masses are much heavier than the gravitino
mass, $m_\phi\,\sim\, 8\pi^2 m_{3/2}$ or even heavier, which might
be considered as an attractive feature in view of the cosmological
moduli problem. However such heavy moduli still cause a cosmological
difficulty, the moduli-induced gravitino problem, producing too many
LSPs which would overclose the Universe if the LSP is given by the
MSSM neutralino. Another potential difficulty of sequestered SUSY
breaking scenario is that it is not straightforward to get a weak
scale size of the Higgs $\mu$ and $B$ parameters.

An attractive way to get a weak scale size of $\mu$ and $B$ in heavy
gravitino scenario is to generate them by a singlet flat direction
which is  stabilized by radiative effects associated with SUSY
breaking. Unless one assumes an unusual type of initial condition,
such a flat direction generically triggers a late thermal inflation.
Thermal inflation would immediately solve the moduli-induced
gravitino problem, however it requires a baryogenesis mechanism to
work after thermal inflation is over.

In this paper, we have presented a model of thermal inflation in
heavy gravitino scenario, which successfully incorporates  the
Affleck-Dine leptogenesis while producing a correct amount of relic
dark matter density. The model involves two singlet flat directions
stabilized by radiative effects, one direction that triggers thermal
inflation and generates a weak scale size of $\mu$ and $B$, and the
other direction that generates the scale of spontaneous lepton
number violation. The dark matter is provided by the lightest
flatino which might be identified as the axino if the model is
assumed to have a $U(1)_{PQ}$ symmetry to solve the strong CP
problem. The collider signal of the model highly depends on the
flaton vacuum values $\langle \varphi\rangle$ ($\varphi=X, S$) which
determine the rate of the decay of the lightest sparticle in the
MSSM to the lighter flatinos. If $\langle \varphi\rangle$ are well
above the intermediate scale $\sim 10^{10}$ GeV, the collider
phenomenology of the model would be almost the same as the
conventional MSSM. For $\langle \varphi\rangle\sim 10^{10}$ GeV or
lower, the model can give a distinct signal associated with the
decay of the lightest sparticle in the MSSM \cite{Nakamura:2008ey},
and this will be the subject of future work \cite{preparation}.

\vskip 1cm {\bf Acknowledgement}

We thank E. D. Stewart for useful discussions, and also D. Lyth for
a comment concerning the AD baryogenesis in gauge mediation.
KC and CSS are supported by the KRF grants funded by the Korean
Government (KRF-2007-341-C00010 and KRF-2008-314-C00064), KOSEF
grant funded by the Korean Government (No. 2009-0080844), and the
BK21 project by the Korean Government. WIP is supported by the Korea
Research Foundation (KRF) grant funded by the Korea government
(MEST) (No. 2009-0077503).


\vskip 1cm

\begin{thebibliography}{99}


\bibitem{Nilles:1983ge}
  H.~P.~Nilles,
  Phys.\ Rept.\  {\bf 110}, 1 (1984);
  H.~E.~Haber and G.~L.~Kane,
  Phys.\ Rept.\  {\bf 117}, 75 (1985);
  D.~J.~H.~Chung, L.~L.~Everett, G.~L.~Kane, S.~F.~King, J.~D.~Lykken and L.~T.~Wang,
  Phys.\ Rept.\  {\bf 407}, 1 (2005)
  [arXiv:hep-ph/0312378].


\bibitem{Pagels:1981ke}
  H.~Pagels and J.~R.~Primack,
  Phys.\ Rev.\ Lett.\  {\bf 48}, 223 (1982);
  S.~Weinberg,
  Phys.\ Rev.\ Lett.\  {\bf 48}, 1303 (1982);
  L.~M.~Krauss,
  Nucl.\ Phys.\  B {\bf 227}, 556 (1983).





\bibitem{Kachru:2003aw}
 S.~B.~Giddings, S.~Kachru and J.~Polchinski,
  Phys.\ Rev.\  D {\bf 66}, 106006 (2002)
  [arXiv:hep-th/0105097];
  S.~Kachru, R.~Kallosh, A.~Linde and S.~P.~Trivedi,
  Phys.\ Rev.\  D {\bf 68}, 046005 (2003)
  [arXiv:hep-th/0301240].



\bibitem{warped_sequestering}
  S.~Kachru, L.~McAllister and R.~Sundrum,
  JHEP {\bf 0710}, 013 (2007)
  [arXiv:hep-th/0703105];
  M.~Son and R.~Sundrum,
  JHEP {\bf 0808}, 004 (2008)
  [arXiv:0801.4789 [hep-th]].


\bibitem{Choi:2006bh}
  K.~Choi and K.~S.~Jeong,
  JHEP {\bf 0608}, 007 (2006)
  [arXiv:hep-th/0605108].


\bibitem{rs}
L.~Randall and R.~Sundrum,
  Nucl.\ Phys.\  B {\bf 557}, 79 (1999)
  [arXiv:hep-th/9810155].

\bibitem{gravity.mediation}
H. P. Nilles, Phys. Lett. B{\bf 115}, 193 (1982); A. H. Chamseddine,
R. L. Arnowitt and P. Nath, Phys. Rev. Lett. {\bf 49}, 970 (1982);
R. Barbieri, S. Ferrara and C. A. Savoy,  Phys. Lett. B{\bf 119},
343 (1982); H. P. Nilles, M. Srednicki and D. Wyler, Phys. Lett.
B{\bf 120}, 346 (1983); L. J. Hall, J. D. Lykken and S. Weinberg,
Phys. Rev. D{\bf 27},2359 (1983);
%
  N.~Ohta,
  Prog.\ Theor.\ Phys.\  {\bf 70}, 542 (1983).



\bibitem{Choi:2004sx}
  K.~Choi, A.~Falkowski, H.~P.~Nilles, M.~Olechowski and S.~Pokorski,
  JHEP {\bf 0411}, 076 (2004)
  [arXiv:hep-th/0411066];
  K.~Choi, A.~Falkowski, H.~P.~Nilles and M.~Olechowski,
  Nucl.\ Phys.\  B {\bf 718}, 113 (2005)
  [arXiv:hep-th/0503216].


\bibitem{Endo:2006zj}
  M.~Endo, K.~Hamaguchi and F.~Takahashi,
  Phys.\ Rev.\ Lett.\  {\bf 96}, 211301 (2006)
  [arXiv:hep-ph/0602061];
  S.~Nakamura and M.~Yamaguchi,
  Phys.\ Lett.\  B {\bf 638}, 389 (2006)
  [arXiv:hep-ph/0602081].




\bibitem{Nakamura:2008ey}
  S.~Nakamura, K.~i.~Okumura and M.~Yamaguchi,
  Phys.\ Rev.\  D {\bf 77}, 115027 (2008)
  [arXiv:0803.3725 [hep-ph]].


\bibitem{Kim:2008hd}
  For a review, see
  J.~E.~Kim,
  Phys.\ Rept.\  {\bf 150}, 1 (1987);
  H.~Y.~Cheng,
  Phys.\ Rept.\  {\bf 158}, 1 (1988);
  J.~E.~Kim and G.~Carosi,
  arXiv:0807.3125 [hep-ph].


\bibitem{Covi:1999ty}
  L.~Covi, J.~E.~Kim and L.~Roszkowski,
  Phys.\ Rev.\ Lett.\  {\bf 82}, 4180 (1999)
  [arXiv:hep-ph/9905212];
  L.~Covi, H.~B.~Kim, J.~E.~Kim and L.~Roszkowski,
  JHEP {\bf 0105}, 033 (2001)
  [arXiv:hep-ph/0101009].



\bibitem{Lyth:1995hj}
  D.~H.~Lyth and E.~D.~Stewart,
  Phys.\ Rev.\ Lett.\  {\bf 75}, 201 (1995)
  [arXiv:hep-ph/9502417];
  Phys.\ Rev.\  D {\bf 53}, 1784 (1996)
  [arXiv:hep-ph/9510204].


\bibitem{de Gouvea:1997tn}
  A.~de Gouvea, T.~Moroi and H.~Murayama,
  Phys.\ Rev.\  D {\bf 56}, 1281 (1997)
  [arXiv:hep-ph/9701244].


\bibitem{Affleck:1984fy}
  I.~Affleck and M.~Dine,
  Nucl.\ Phys.\  B {\bf 249}, 361 (1985);
  M.~Dine, L.~Randall and S.~D.~Thomas,
  Nucl.\ Phys.\  B {\bf 458}, 291 (1996)
  [arXiv:hep-ph/9507453].


\bibitem{Jeong:2004hy}
  D.~h.~Jeong, K.~Kadota, W.~I.~Park and E.~D.~Stewart,
  JHEP {\bf 0411}, 046 (2004)
  [arXiv:hep-ph/0406136].

\bibitem{Felder:2007iz}
  G.~N.~Felder, H.~Kim, W.~I.~Park and E.~D.~Stewart,
  JCAP {\bf 0706}, 005 (2007)
  [arXiv:hep-ph/0703275].


\bibitem{Kim:2008yu}
  S.~Kim, W.~I.~Park and E.~D.~Stewart,
  JHEP {\bf 0901}, 015 (2009)
  [arXiv:0807.3607 [hep-ph]].

\bibitem{anomaly.mediation}
  L.~Randall and R.~Sundrum,
  Nucl.\ Phys.\  B {\bf 557}, 79 (1999)
  [arXiv:hep-th/9810155];
  G.~F.~Giudice, M.~A.~Luty, H.~Murayama and R.~Rattazzi,
  JHEP {\bf 9812}, 027 (1998)
  [arXiv:hep-ph/9810442];
  J.~A.~Bagger, T.~Moroi and E.~Poppitz,
  JHEP {\bf 0004}, 009 (2000)
  [arXiv:hep-th/9911029].

\bibitem{dilaton.mediation}
  V.~S.~Kaplunovsky and J.~Louis,
  Phys.\ Lett.\  B {\bf 306}, 269 (1993)
  [arXiv:hep-th/9303040];
  A.~Brignole, L.~E.~Ibanez and C.~Munoz,
  Nucl.\ Phys.\  B {\bf 422}, 125 (1994)
  [Erratum-ibid.\  B {\bf 436}, 747 (1995)]
  [arXiv:hep-ph/9308271].


\bibitem{Choi:2008hn}
  K.~Choi, K.~S.~Jeong and K.~I.~Okumura,
  JHEP {\bf 0807}, 047 (2008)
  [arXiv:0804.4283 [hep-ph]].

\bibitem{Endo:2005uy}
  M.~Endo, M.~Yamaguchi and K.~Yoshioka,
  Phys.\ Rev.\  D {\bf 72}, 015004 (2005)
  [arXiv:hep-ph/0504036].


\bibitem{Choi:2005uz}
  K.~Choi, K.~S.~Jeong and K.~i.~Okumura,
  JHEP {\bf 0509}, 039 (2005)
  [arXiv:hep-ph/0504037].


\bibitem{moroi_randall}
  T.~Moroi and L.~Randall,
  Nucl.\ Phys.\  B {\bf 570}, 455 (2000)
  [arXiv:hep-ph/9906527].


\bibitem{kane_etal}
  B.~S.~Acharya, P.~Kumar, K.~Bobkov, G.~Kane, J.~Shao and S.~Watson,
  JHEP {\bf 0806}, 064 (2008)
  [arXiv:0804.0863 [hep-ph]].


\bibitem{Rattazzi:2000}
  A. Pomarol and R. Rattazzi, JHEP {\bf 9905}, 013 (1999)
  [arXiv:hep-ph/9903448];
  R.~Rattazzi, A.~Strumia and J.~D.~Wells,
  Nucl.\ Phys.\  B {\bf 576}, 3 (2000)
  [arXiv:hep-ph/9912390].


\bibitem{Chun:2000jx}
  E.~J.~Chun, H.~B.~Kim and D.~H.~Lyth,
  Phys.\ Rev.\  D {\bf 62}, 125001 (2000)
  [arXiv:hep-ph/0008139];
  E.~J.~Chun, H.~B.~Kim, K.~Kohri and D.~H.~Lyth,
  JHEP {\bf 0803}, 061 (2008)
  [arXiv:0801.4108 [hep-ph]].



\bibitem{Choi:1996vz}
  K.~Choi, E.~J.~Chun and J.~E.~Kim,
  Phys.\ Lett.\  B {\bf 403} (1997) 209
  [arXiv:hep-ph/9608222].


\bibitem{seesaw}
T.Yanagida, in $Proceedings$ $of~the$ $Workshop$ $on$ $Unified$
$Theories$ $and$ $Baryon$ $Number$ $in~the$ $Universe$, eds. O.
Sawada and A. Sugamoto (KEK, Tsukuba, 1979); M. Gell-Mann, P. Ramond
and R. Slansky, in $Supergravity$, eds. P. van Nieuwenhuizen and D.
Freedman (North Holland, Amsterdam, 1979);
 R.N. Mohapatra and G.
Senjanovic, Phys. Rev. Lett. 44 (1980) 912.

\bibitem{Everett:2008qy}
  L.~L.~Everett, I.~W.~Kim, P.~Ouyang and K.~M.~Zurek,
  Phys.\ Rev.\ Lett.\  {\bf 101}, 101803 (2008)
  [arXiv:0804.0592 [hep-ph]];
  L.~L.~Everett, I.~W.~Kim, P.~Ouyang and K.~M.~Zurek,
  JHEP {\bf 0808}, 102 (2008)
  [arXiv:0806.2330 [hep-ph]];
  K.~Choi, K.~S.~Jeong, S.~Nakamura, K.~I.~Okumura and M.~Yamaguchi,
  JHEP {\bf 0904}, 107 (2009)
  [arXiv:0901.0052 [hep-ph]];
  M.~Holmes and B.~D.~Nelson,
  JCAP {\bf 0907}, 019 (2009)
  [arXiv:0905.0674 [hep-ph]].


\bibitem{Comelli:1996vm}
  D.~Comelli and J.~R.~Espinosa,
  Phys.\ Rev.\  D {\bf 55}, 6253 (1997)
  [arXiv:hep-ph/9606438].


\bibitem{Yao:2006px}
  W.~M.~Yao {\it et al.}  [Particle Data Group],
  J.\ Phys.\ G {\bf 33} (2006) 1.

\bibitem{Kleban:2007jd}
  M.~Kleban, K.~Sigurdson and I.~Swanson,
  JCAP {\bf 0708}, 009 (2007)
  [arXiv:hep-th/0703215].

\bibitem{Casas:1995pd}
  J.~A.~Casas, A.~Lleyda and C.~Munoz,
  Nucl.\ Phys.\  B {\bf 471}, 3 (1996)
  [arXiv:hep-ph/9507294].


\bibitem{Bashinsky:2003tk}
  S.~Bashinsky and U.~Seljak,
  Phys.\ Rev.\  D {\bf 69}, 083002 (2004)
  [arXiv:astro-ph/0310198].


\bibitem{Hannestad:2003ye}
  S.~Hannestad and G.~Raffelt,
  JCAP {\bf 0404}, 008 (2004)
  [arXiv:hep-ph/0312154].



\bibitem{Crotty:2004gm}
  P.~Crotty, J.~Lesgourgues and S.~Pastor,
  Phys.\ Rev.\  D {\bf 69}, 123007 (2004)
  [arXiv:hep-ph/0402049].


\bibitem{Hannestad:2008js}
  S.~Hannestad, A.~Mirizzi, G.~G.~Raffelt and Y.~Y.~Y.~Wong,
  JCAP {\bf 0804}, 019 (2008)
  [arXiv:0803.1585 [astro-ph]].


\bibitem{Amsler:2008zzb}
  C.~Amsler {\it et al.}  [Particle Data Group],
  Phys.\ Lett.\  B {\bf 667}, 1 (2008).


\bibitem{Jedamzik:2005sx}
  K.~Jedamzik, M.~Lemoine and G.~Moultaka,
  JCAP {\bf 0607}, 010 (2006)
  [arXiv:astro-ph/0508141].


\bibitem{Dunkley:2008ie}
  J.~Dunkley {\it et al.}  [WMAP Collaboration],
  Astrophys.\ J.\ Suppl.\  {\bf 180}, 306 (2009)
  [arXiv:0803.0586 [astro-ph]].


\bibitem{KT}
E.~W.~Kolb and M.~S.~Turner, ``The early universe'',
(Addison-Wesley, 1990).


\bibitem{preparation}
K. Choi, K. S. Jeong, C. B. Park and S. H. Im, in preparation.

\end{thebibliography}
\end{document}